\documentclass[twocolumn,aps,prd,nofootinbib,showpacs,superscriptaddress,floatfix]{revtex4-2}
\usepackage{amsmath,amsfonts,amssymb}
\usepackage{graphicx}
\usepackage[usenames]{color}
\usepackage[letterpaper,margin=0.75in]{geometry}
\usepackage{multirow}
\usepackage{array}
\usepackage[citecolor=blue,pdfa=true,urlcolor=magenta,colorlinks=True]{hyperref}

\usepackage{ifthen}

\newcommand{\sorthelp}[1]{}
\newcommand{\aap}{Astron. Astrophys.}
\newcommand{\jcap}{J. Cosmol. Astropart. Phys.}
\newcommand{\apjs}{Astrophys. J. Suppl. Ser.}
\newcommand{\mnras}{Mon. Not. R. Astron. Soc.}

\def\beq{\begin{equation}}
\def\eeq{\end{equation}}
\def\bea{\setlength\arraycolsep{1.4pt}\begin{eqnarray}}
\def\eea{\end{eqnarray}}
\def\bit{\begin{itemize}}
\def\eit{\end{itemize}}

\def\eq{Eq.~}
\def\eqs{Eqs.~}

\def\ld{\left}
\def\rd{\right}

\def\fr{\frac}

\def\bb{\widetilde{\omega}_{\rm b}}
\def\bc{\widetilde{\omega}_{\rm c}}
\def\bht{\Omega_{\rm b} h^2}
\def\cht{\Omega_{\rm c} h^2}
\def\lc{\Lambda{\rm CDM}}

\def\mnras{MNRAS}
\def\apjs{ApJS}
\def\apj{ApJ}
\def\aap{A\&A}
\def\apjl{ApJ}
\def\jcap{JCAP}
\def\prd{Phys. Rev. D}
\def\apss{Ap\&SS} 

\def\nhat{{\hat{\mathbf{n}}}}
\def\mhat{{\hat{\mathbf{m}}}}

\begin{document}

\title{The role of \texorpdfstring{\boldmath{$T_0$}}{T0} in CMB anisotropy measurements}

\author{Yunfei Wen}\email{y52wen@uwaterloo.ca}
\affiliation{Department of Physics \& Astronomy\\
University of British Columbia, Vancouver, BC, V6T 1Z1  Canada} 
\affiliation{Department of Applied Mathematics\\
University of Waterloo, Waterloo, ON, N2L 3G1 Canada}

\author{Douglas Scott}\email{dscott@phas.ubc.ca}
\affiliation{Department of Physics \&
Astronomy\\
University of British Columbia, Vancouver, BC, V6T 1Z1 Canada}

\author{Raelyn Sullivan}\email{rsullivan@phas.ubc.ca}
\affiliation{Department of Physics \& Astronomy\\
University of British Columbia, Vancouver, BC, V6T 1Z1  Canada}

\author{J. P. Zibin}\email{zibin@cosm.ca}
\noaffiliation

\date{\today}

\begin{abstract}
The quantity $T_0$, the cosmic microwave background (CMB) monopole, is an often neglected seventh parameter of the standard cosmological model. As well as its variation affecting the physics of the CMB, the measurement of $T_0$ is also used to calibrate the anisotropies, via the orbital dipole. We point out that it is easy to misestimate the effect of $T_0$ because the CMB anisotropies are conventionally provided in temperature units.  In fact the anisotropies are most naturally described as dimensionless and we argue for restoring the convention of working with $\Delta T/T$ rather than $\Delta T$.  As a free cosmological parameter, $T_0$ most naturally only impacts the CMB power spectra through late-time effects.  Thus if we ignore the {\it COBE}-FIRAS measurement, current CMB data only weakly constrain $T_0$.  Even ideal future CMB data can at best provide a percent-level constraint on $T_0$, although adding large-scale structure data will lead to further improvement.  The FIRAS measurement is so precise that its uncertainty negligibly affects most, but not all, cosmological parameter inferences for current CMB experiments. However, if we eventually want to extract all available information from CMB power spectra measured to multipoles $\ell\simeq5000$, then we will need a better determination of $T_0$ than is currently available.
\end{abstract}

\maketitle

\section{\label{sec:level1}Introduction}
Experimental results in the past couple of decades have established $\Lambda$ cold dark matter ($\Lambda$CDM) as the standard model of cosmology. This era of precision cosmology has been largely driven by studies of the cosmic microwave background (CMB), which demonstrate that a 6-parameter model provides a good fit to the data. The basic parameter set is often written as $\{\Omega_{\rm b}h^2, \Omega_{\rm c}h^2, \theta_{\ast}, A_{\rm s}, n_{\rm s}, \tau\}$, which has been tightly constrained by \textit{Planck}~\cite{2013Planckparameter,2015Planckparameter,2018Planckparameter} and other experiments~\cite{WMAP9yearCosmologicalParameter, 2020ACTDR4}.
Despite this being described as the 6-parameter $\Lambda$CDM model, there is in fact a seventh parameter, namely $T_0$, the present-day temperature of the photon background, or the monopole term in the spherical-harmonic expansion of the CMB sky. This additional parameter is usually neglected because it is so well measured that it can be regarded as effectively fixed.

In current cosmological analyses, $T_0$ is given by the precise measurements available from a combination of data from \textit{COBE}-FIRAS~\cite{FIRAS1996} and other experiments~\cite{FIRAStemp},
yielding
\begin{equation}
T_{0,{\rm F}}=(2.7255\pm0.0006) \,{\rm K},
\label{eq: FIRAS_Error}
\end{equation}
which has an uncertainty at the level of 0.02\,\%; for simplicity we will refer to this as the ``FIRAS'' temperature. The \textit{Planck} 2018 results give values of $\theta_{\ast}$ and $\Omega_{\rm b}h^2$ that are constrained to approximately the $0.03\,\%$ and $0.67\,\%$ levels, respectively~\cite{2018Planckparameter}, which begin to approach the precision of $T_{0,{\rm F}}$. In the future, with continuously improving measurements of higher multipoles of CMB power spectra from ground-based experiments~\cite{2014SPT-3G,2014ACTPol,2016CMBS4}, and the possibility of ambitious space-based experiments~\cite{2018CORE}, it will be possible to approach a cosmic-variance-limited (CVL) measurement for power spectra to high multipoles, particularly for polarization, where there are not expected to be significant small-scale foreground signals.  With correspondingly dramatic improvements in the derived cosmological parameters, in this paper we assess whether it is reasonable to continue treating temperature as a constant in the cosmological analysis or whether we will eventually need a better measurement of the CMB monopole in order to fully exploit the constraining power of future experimental data. 

\begin{table*}[htbp!]
\begin{tabular}{p{0.155\textwidth}<{\centering}p{0.155\textwidth}<{\centering}p{0.155\textwidth}<{\centering}p{0.155\textwidth}<{\centering}p{0.155\textwidth}<{\centering}p{0.155\textwidth}<{\centering}}
\hline
\hline
\noalign{\vskip 2pt}
 $\Omega_{\rm b} h^2$ & $\Omega_{\rm c} h^2$ & $100\theta_{\ast}$ & $10^9A_{\rm s}$ & $n_{\rm s}$ & $\tau$\\
\hline
\noalign{\vskip 1pt}
0.02237$\pm$0.00015 & 0.1200$\pm$0.0012 & 1.04110 $\pm$0.00031 & 2.100$\pm$0.030 & 0.9649$\pm$0.0042 & 0.0544$\pm$0.0073\\
\hline
\end{tabular}
\caption{68\,\% confidence intervals for the 6-parameter base-$\Lambda$CDM model from the 2018 {\em Planck} TT,TE,EE+lowE+lensing likelihood~\cite{2018Plancklikelihood,2018Planckparameter}. For all the analyses in this paper, we choose the central values of the above intervals as the fiducial model, and we adopt the above error bars as the standard uncertainties for reference.  Note that if we use the FIRAS prior in the 7-parameter $\lc{+}T_0$ model, we obtain essentially identical results.
\label{tab:Planck2018TTTEEELensingResult}
}
\end{table*}

Once we start to consider $T_0$ as a variable, the current way of presenting CMB data in temperature units needs to be re-examined. This is because, besides being a cosmological parameter, $T_0$ is also used for the calibration and presentation of CMB data. The theoretical power spectra $C_{\ell}$ calculated by cosmological codes (e.g., \texttt{CAMB}~\cite{CAMB} or \texttt{CLASS}~\cite{CLASSII}) are frequently presented in temperature units [${\rm \mu K^2}$]. This is despite the fact that from the time of the earliest discussions of CMB anisotropies, theorists tended to frame their calculations in terms of fractional perturbations, i.e., $\Delta T/T$.  This can be seen in early papers by Peebles~\cite{Peebles1965}, Sachs \& Wolfe~\cite{SW1967}, Silk~\cite{Silk1968}, and Sunyaev \& Zeldovich~\cite{SZ1970}. Conversely, the results of anisotropy experiments, including both the measured Solar dipole and the measured CMB power spectra, are usually expressed in temperature units.  This may seem natural, given that a Planck distribution is fully described by a temperature; however, as we discuss later, the calibration of measurements also has to be considered. As measurements became more precise and the annual variation of the dipole started being used as the primary calibration source, the fact that experiments are directly measuring a dimensionless quantity has become more obscured.

To understand $T_0$ as a parameter, it is necessary to disentangle the role of $T_0$ in the calibration process from $T_0$ in the physics. The situation would be more straightforward if $\Delta T/T$ units were used in presenting data as well as for discussing theoretical predictions. Our purpose here is to clarify these issues, in order to ultimately decide whether variations of $T_0$ within a 7-parameter model actually matter.

In this paper we first review the issue of CMB calibration and discuss the problem of using temperature units for CMB data in Sec.~\ref{sec:calibration}. In Sec.~\ref{T0inCosmo}, we present a pedagogical argument for the theoretical role of $T_0$ in a 7-parameter $\lc$ model (primarily building on the work of Ref.~\cite{T0H0}), and provide constraints on $T_0$ from current and future CMB power spectra and other data.  Next, in Sec.~\ref{ImpactFIRASerror} we analyse the impact of the current $T_0$ uncertainty (from the FIRAS measurement) on cosmological parameter constraints in current and future CMB experiments. We conclude in Sec.~\ref{conclusions}.

Throughout the paper, we use a Fisher-matrix formalism to forecast parameter constraints from CMB data (also including measurements of the baryon acoustic oscillations).  The method used and conditions set for our calculations are described in detail in Appendix~\ref{sec:method}. Since we use several different symbols to refer to various definitions of the background temperature, we give a table of definitions for reference in Appendix~\ref{sec:tempdefs}.  Lastly, we also discuss the relationship between $T_0$ and inhomogeneity in Appendix~\ref{sec:inhomogeneous}.  Throughout we adopt a fiducial flat $\lc$ model with the parameter values in Table~\ref{tab:Planck2018TTTEEELensingResult}, from \textit{Planck}~\cite{2018Plancklikelihood, 2018Planckparameter}, which represents the smallest uncertainties we currently have for parameters in $\lc$ from CMB data alone. Except where explicitly stated otherwise, all the parameter uncertainties are given as $\pm1\,\sigma$, which corresponds to the $68\,\%$ confidence interval for a Gaussian distribution.

\section{Calibration and units}\label{sec:calibration}
\subsection{Motivation for dimensionless fluctuations}\label{calibration}

The CMB temperature anisotropy power spectrum $C_\ell$ [which is related to the multipole-scaled quantity $D_\ell=\ell(\ell+1)C_\ell/(2\pi$)] is usually defined as the covariance of the coefficients of the spherical harmonic expansion of the dimensionless quantity $\Delta T(\hat{\mathbf{n}})/T$, where $\hat{\mathbf{n}}$ is the direction on the sky. The predictions of theoretical models for the power spectra are also most directly obtained as dimensionless quantities. Indeed, the dimensionless anisotropy power is directly related to the dimensionless amplitude of primordial curvature perturbations, $A_{\rm s}$, and for the primary anisotropies the dimensionless power (unlike the power in temperature units) satisfies the approximate scaling relation~\cite{ZMS2007}
\beq
\ell'^2 C_{\ell'}(t') \simeq \ell^2 C_\ell(t)
\eeq
during the evolution of the Universe, where
\beq
\ell' \equiv \ell \frac{D_{\rm A}(t')}{D_{\rm A}(t)},
\eeq
for angular diameter distance $D_{\rm A}(t)$ from recombination to the observer at time $t$. However, in the current convention the temperature and polarization anisotropy power spectra are given in temperature units (usually $\mu{\rm K}^2$) for measured results from CMB experiments and, usually, also for the results of theoretical calculations.  On 
the other hand the lensing reconstruction spectrum, $C_\ell^{\phi\phi}$, is always provided in dimensionless form.

While conventions can be difficult to change, we advocate here for the use of dimensionless quantities exclusively, in both theoretical and experimental studies. Apart from their simpler theoretical properties described above, dimensionless quantities are also advantageous from the standpoint of calibration. All CMB anisotropy experiments use differential measurements, and so they are not directly sensitive to the value of the temperature monopole (but see below for further discussion on this point). In a typical CMB experiment, the detectors measure, in volts, differences between the brightness of positions on the sky. These data then need to be calibrated. The most precise calibration is the annual time-variation of the dipole, often referred to as the ``orbital dipole''. This is known to extremely high precision (in velocity, or $v/c$ or $\Delta T/T$ units) because of the complete knowledge of the satellite's orbital motion~\cite{2015HFIcalibration}. This calibration then allows the annually-averaged dipole (usually called the ``Solar dipole'') to be extracted from the data, also in dimensionless units.  It should be clear that this argument for the $\ell=1$ mode applies equally well to all the other multipole coefficients of the measured CMB sky, so they too are fundamentally measured in velocity, or dimensionless, units.

To help elucidate the advantages of dimensionless quantities, it will be useful 
to consider the calibration of CMB anisotropy experiments in more detail; here we focus on the {\it Planck\/} satellite, but we expect the same discussion to apply generally. The approaches for calibration of data from {\it Planck\/} are described in great detail in a series of papers \cite{PreLaunchHFIcalibration,2013LFIcalibration,2013HFIcalibration,2015LFIcalibration,2015HFIcalibration,2018LFIcalibration,2018HFIcalibration}. For clarity of the treatment, we ignore foregrounds for the moment and assume that only CMB anisotropy is being observed. {\it Planck\/} measures a signal (a detector voltage) that (ignoring beam effects, which could easily be added) is proportional to the intensity difference between two directions on the sky,
\beq
\Delta I_\nu(\nu,\nhat) \equiv I_\nu(\nu,\nhat) - I_\nu(\nu,\mhat)
\simeq \ld.\fr{dI_\nu}{dT}\rd|_\mhat \Delta T(\nhat)
\label{delInu}
\eeq
at first order in $\Delta T(\nhat)$, where $\mhat$ is a (fixed) reference direction, 
with $T(\mhat) \equiv T_0$ and $\Delta T(\nhat) \equiv T(\nhat) - T_0$, and 
$I_\nu(\nu,\nhat)$ is the Planck intensity distribution.  When {\em Planck} uses 
the orbital dipole for calibration, the amplitude of that dipole, $v_{\rm od}/c 
= \Delta T_{\rm od}/T_0$, is known with very high precision.  Therefore, according 
to \eq(\ref{delInu}), at first order in $\Delta T(\nhat)$ we have
\beq
\fr{v_{\rm od}}{c} = \fr{\Delta I_{\nu,{\rm od}}(\nu)}{T_0\,dI_\nu/dT|_\mhat}
   = \fr{e^{x_0} - 1}{x_0e^{x_0}}\fr{\Delta I_{\nu,{\rm od}}(\nu)}{I_\nu(\nu,\mhat)},
\label{vodc}
\eeq
where $\Delta I_{\nu,{\rm od}}(\nu)$ is the intensity difference corresponding 
to the orbital dipole amplitude and $x_0(\nu) \equiv h\nu/(kT_0)$.  Therefore 
the orbital dipole calibration gives directly, at the same precision as we know 
$v_{\rm od}/c$, the quantity on the right-hand side of \eq(\ref{vodc}), and {\it not\/} the 
intensity difference $\Delta I_{\nu,{\rm od}}/I_\nu$.  It is only when we 
also know $T_0$ (and hence $x_0$) precisely that the calibration also gives us 
$\Delta I_{\nu,{\rm od}}/I_\nu$.

Next, when {\em Planck} measures a CMB fluctuation in some direction $\nhat$, 
combining \eqs(\ref{delInu}) and (\ref{vodc}) gives, at first order,
\beq
\fr{\Delta T(\nhat)}{T_0}
   = \fr{\Delta I_\nu(\nu,\nhat)}{\Delta I_{\nu,{\rm od}}(\nu)}\fr{v_{\rm od}}{c}.
\label{obsdTCMB}
\eeq
Thus we can determine 
the (dimensionless) temperature fluctuation knowing only the ratio 
$\Delta I_\nu(\nu,\nhat)/\Delta I_{\nu,{\rm od}}(\nu)$, which is equal to the (directly 
measured) corresponding detector voltage ratio.  In particular, we do not need to 
know $T_0$, so if the $T_0$ value we adopted was completely wrong we would 
still obtain the correct dimensionless $\Delta T(\nhat)/T_0$ for the fluctuations.

It may be worth adding that for the two highest frequency channels of {\em Planck}, where the CMB anisotropies do not dominate even at high Galactic latitudes, the calibration procedure instead uses the brightness of planets.  This is most naturally done in intensity units, i.e., $\Delta I_\nu$ rather than $\Delta T/T$.  However, the precision of the planetary calibration process only reaches the
1\,\% level under the most optimistic set of assumptions \cite{Bertincourt} and hence has no bearing on the details of the much more precise $T_0$ calibration that we are discussing here.

When we include foregrounds, the total signal is often written in terms of the 
Rayleigh-Jeans temperature fluctuation, explicitly (see, e.g., \cite{eriksenetal06})
\beq
\Delta T_{\rm RJ}(\nu,\nhat) = \fr{x_0^2 e^{x_0}}{(e^{x_0} - 1)^2}\Delta T_{\rm CMB}(\nhat)
   + \sum_i A_i(\nhat) F_i(\nu,\nhat,\theta),
\label{DelTRJ}
\eeq
where the sum is over foreground components with amplitudes $A_i$ and frequency 
dependence $F_i$, $\theta$ is a set of foreground parameters, and the 
Rayleigh-Jeans temperature fluctuation is given by
\beq
\Delta T_{\rm RJ}(\nu,\nhat) = \fr{\Delta I_\nu(\nu,\nhat)c^2}{2k\nu^2}.
\label{delRJdef}
\eeq
The relation corresponding to \eq(\ref{obsdTCMB}) for the Rayleigh-Jeans temperature 
fluctuation is
\beq
\fr{(e^{x_0} - 1)^2}{x_0^2 e^{x_0}}\fr{\Delta T_{\rm RJ}(\nu,\nhat)}{T_0}
   = \fr{\Delta I_\nu(\nu,\nhat)}{\Delta I_{\nu,{\rm od}}(\nu)}\fr{v_{\rm od}}{c}.
\label{scaledRJ}
\eeq
Again, the quantity on the right-hand side of this equation is directly measured when using 
the orbital dipole for calibration, and is independent of $T_0$; however, now we need 
$T_0$ if we wish to determine the Rayleigh-Jeans fluctuation.  We can nevertheless write 
\eq(\ref{DelTRJ}) in a form that is dimensionless and independent of the value 
of $T_0$ as
\begin{align}
\fr{(e^{x_0} - 1)^2}{x_0^2 e^{x_0}}\fr{\Delta T_{\rm RJ}(\nu,\nhat)}{T_0}
   &= \fr{\Delta T_{\rm CMB}(\nhat)}{T_0}\label{measFG}\\
   &+ \fr{(e^{x_0} - 1)^2}{x_0^2 e^{x_0}}\sum_i \fr{A_i(\nhat)}{T_0}F_i(\nu,\nhat,\theta).\nonumber
\end{align}
Here the foreground amplitude parameters will have units of temperature, so indeed 
this expression is dimensionless.  Again, we stress that the quantity on the 
left-hand side is {\em directly} measured in terms of the calibration, {\em independently of 
$T_0$}, according to \eq(\ref{scaledRJ}).

Now finally, if we imagine a hypothetical situation where our adopted value of $T_0$ is wrong, the measured 
left-hand side of \eq(\ref{measFG}) will not change.  However, the $T_0$- (and $\nu$-) 
dependent factors multiplying the foreground sum {\em will} change.  To the extent 
that the foreground parameters are degenerate with this change, we could still 
recover the correct CMB fluctuations; however, the foreground parameters would be 
biased from their true values.  For a large enough error in our adopted $T_0$, we 
expect the shift in $T_0$ to no longer be degenerate with shifts in foreground 
parameters, so we would have a poor fit, and might not correctly recover the CMB anisotropies.  
This is not surprising, considering that using an incorrect value of $T_0$ means 
we might be assuming in error that the CMB dominates (or does not) over a particular 
foreground component in some frequency range.  In principle, to the extent that we are 
confident about which foreground components are important and can place priors on their parameters, \eq(\ref{measFG}) implies that we could 
place a constraint on $T_0$ from the {\em Planck} frequency maps alone.  However, for this to be useful, we would need a very precise first-principles calculation of the foreground parameters.  In practice such parameters will have fairly wide priors and hence any constraint on $T_0$ from matching foregrounds will be quite weak.

To summarize, it is the dimensionless CMB fluctuation $\Delta T/T_0$ that is most directly constrained in terms of the orbital dipole calibration.  Ignoring foregrounds, 
we can determine $\Delta T/T_0$ this way independently of the monopole $T_0$.  
In the presence of foregrounds, errors in $T_0$ may bias the foreground parameters, 
though in practice the uncertainties in those parameters should far exceed the 
level of the FIRAS $T_0$ uncertainty.

\subsection{Conversion to temperature units} \label{unit}

If we do wish to give the amplitude of the Solar dipole and higher multipoles in temperature units, it is necessary to multiply by a monopole value. Here there are two slightly different questions that we can ask.  The first is ``what is the best we can say about $C_\ell$ in temperature units?''  If $C_{\ell}^{\rm [D]}$ is the measured dimensionless spectrum and $C_{\ell}^{\rm [T]}$ is in temperature units,\footnote{Appreciating that it is unorthodox for the units to change the meaning of a variable, here we put the label in square brackets to indicate that we are explicitly referring to a particular choice of units.} the answer is
\begin{equation}
C_\ell^{\rm [T]} = T_0^2C_\ell^{\rm [D]}(T_0,\mathbf{p}), \label{eq:C_l_T_unit}
\end{equation}
where $\mathbf{p}$ represents the remaining cosmological parameters. The quantity $C_{\ell}^{\rm [T]}$ in Eq.~(\ref{eq:C_l_T_unit}) is equivalent to the covariance of the $a_{lm}$s when we directly expand $\Delta T(\hat{\mathbf{n}})$ (instead of $\Delta T(\hat{\mathbf{n}})/T$) in spherical harmonics. Here the uncertainty in the FIRAS-derived value $T_{0,{\rm F}}$ propagates into the uncertainty in $C_\ell^{\rm [T]}$ via the calibration, as it must. Note that, if we wish to compare the measured spectra with model predictions using this conversion method, we must convert both the predicted spectra and the measured spectra (despite the fact that the measured spectra are fundamentally dimensionless in nature) into temperature units, taking account of the uncertainty in $T_0$. In most practical settings, when we are not considering $T_0$ to be a free parameter, consideration of the uncertainty in $T_0$ simply means taking the FIRAS measurement $T_{0,{\rm F}}$ in Eq.~(\ref{eq: FIRAS_Error}).

The second approach to converting the spectra to temperature units is the answer to the question: ``what is the best we can say about the anisotropy power?''  The answer is now
\begin{equation}
C_\ell^{\rm [T]} = T_{\rm c}^2C_\ell^{\rm [D]}(T_0,\mathbf{p}), \label{eq:C_l unit}
\end{equation}
where $T_{\rm c}$ is a fixed calibration temperature that we select in order to convert $C_{\ell}$ from dimensionless to temperature units. Since $T_{\rm c}$ is constant (conventionally the central value of the FIRAS measurement, \eq{\ref{eq: FIRAS_Error}}, hereafter $\overline{T}_{0,\rm F}$), the $T_0$ uncertainty does {\it not\/} propagate into the errors in $C_\ell^{\rm [T]}$, and so we know $C_\ell^{\rm [T]}$ with greater precision than if we had used Eq.~(\ref{eq:C_l_T_unit}). Of course, when using 
Eq.~(\ref{eq:C_l unit}) there is no actual dependence on the experimental value of the monopole in the calibration, since one could always divide out the fixed value of $T_{\rm c}$ that was used and recover the original dimensionless measurement. So it is equivalent, but simpler, to stay dimensionless from the start. Another potential problem with choosing a factor $T_{\rm c}$ is that a future measurement of $T_0$~\cite{PIXIE_cal,PRISM} might give a different central value, complicating comparisons. 

In practice with current experiments, it makes essentially no difference which calibration approach we take as far as the higher multipoles ($\ell \ge 2$) are concerned, since the FIRAS uncertainty is negligible compared to other errors. However, for the Solar dipole, folding in the FIRAS error does make a substantial contribution to the total uncertainty.  In this case, when quoting the results in velocity units the Planck Collaboration papers never include the FIRAS error.  However, the situation is not always consistent when it comes to presenting the uncertainty of the Solar dipole in temperature units and choosing whether to add the FIRAS uncertainty in quadrature~\cite{2015HFIcalibration,2018Planckoverview,2018LFIcalibration,2018HFIcalibration,2020PlanckNPIPE}. Nevertheless, in their latest results~\cite{2020PlanckNPIPE,BeyondPlanck} the collaboration does include the FIRAS uncertainty in the temperature-units dipole. The resulting inflation of uncertainties should be taken into account when comparing different dipole measurements. In summary, {\em Planck} uses both approaches listed above for converting the dipole amplitude to temperature units. While neither approach is wrong (since they provide answers to slightly different questions), we consider that the second approach of choosing fixed $T_{\rm c}$, without combining the FIRAS error, is the more appropriate one. This is because it correctly reflects the dimensionless nature of the dipole measurement and more accurately summarizes our current knowledge of the Solar dipole uncertainty.

\begin{figure}[htbp!]
\centerline{\includegraphics[width=\hsize]{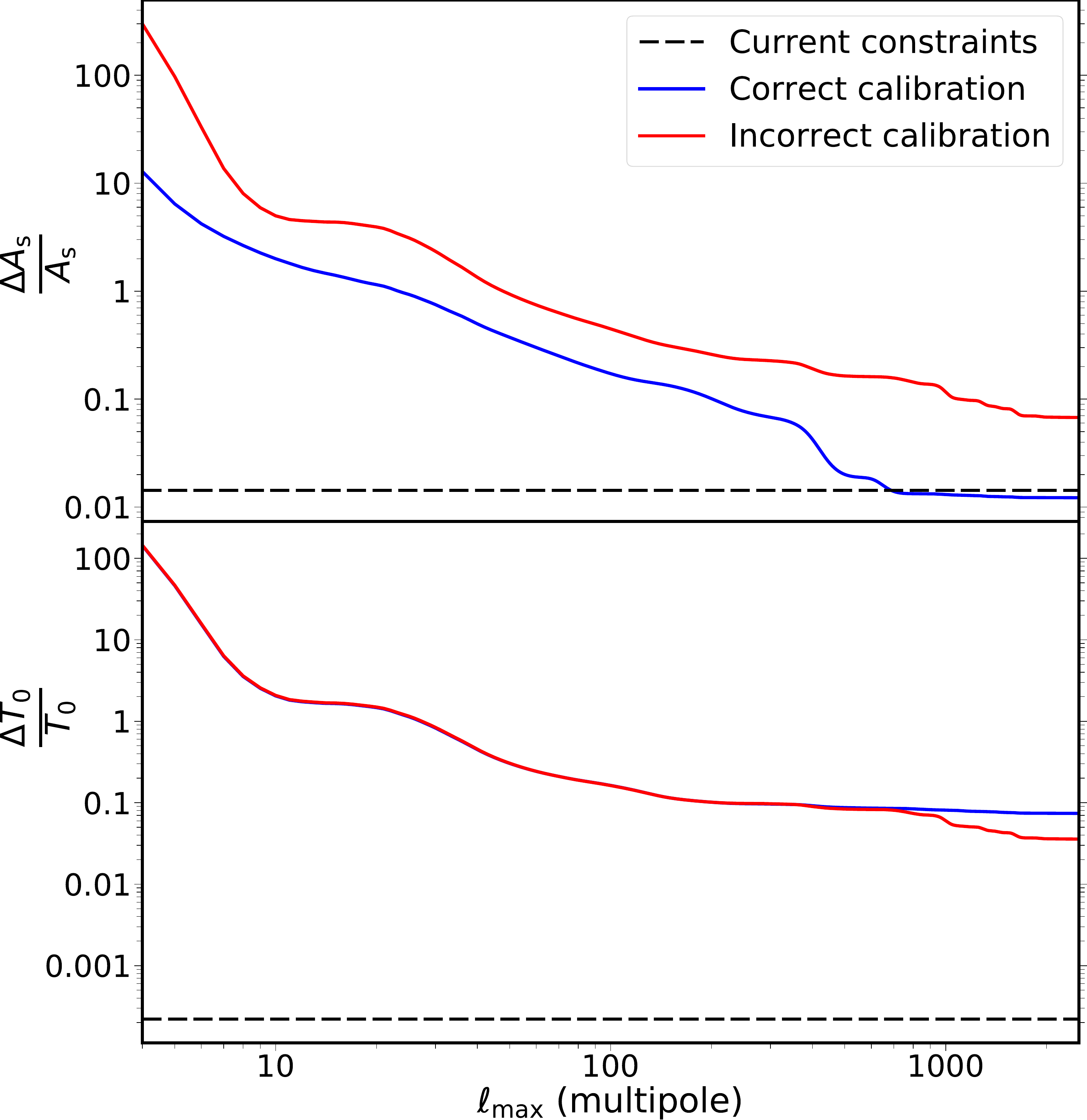}}
\caption{Predicted uncertainties on the power spectrum amplitude parameter $A_{\rm s}$ and the CMB monopole temperature $T_0$ using lensed $TT$, $TE$, and $EE$ power spectra.  Including noise at the level of the \textit{Planck} experiment, we show the relative error estimate for $T_0$ and $A_{\rm s}$ with respect to the maximum $\ell$, with both the correct (\eq{\ref{eq:C_l unit}}) and incorrect (\eq{\ref{eq:C_l_T_unit}}) temperature calibration procedures (as described in Sec.~\ref{sec:calibration}) for the 7-parameter $\lc{+}T_0$ model. The horizontal black dashed lines give the uncertainty in $A_{\rm s}$ derived from the actual 2018 {\em Planck} TT,TE,EE+lensing likelihood and the uncertainty from the FIRAS $T_0$ measurement.}
\label{fig:planck_error_T0}
\end{figure}

Although when comparing the results of higher multipoles in present-day experiments with theoretical calculations it makes essentially no difference whether one uses temperature units or dimensionless quantities, there \textit{are} situations in which this distinction can be crucial. In particular, when considering a 7-parameter $\lc{+}T_0$ model, with variable background temperature, we should calculate the theoretical dimensionless power spectra by varying $T_0$ along with the usual cosmological parameters, but we should \textit{not} include the $T_0$ variation in the calibration of the power spectra. In other words, we should use Eq.~(\ref{eq:C_l unit}) instead of Eq.~(\ref{eq:C_l_T_unit}). If we used Eq.~(\ref{eq:C_l_T_unit}) (and did not similarly scale the calibration of the measured spectra we compare with), the variation of $T_0$ would cause an additional change of the overall amplitude of the power spectra, which would lead to a strong degeneracy between $T_0$ and 
$A_{\rm s}$. In addition, the fact that the lensing spectrum $C_\ell^{\phi\phi}$ is always presented in dimensionless form can lead to an artificially strong lensing constraint~\cite{T0H0}.  As seen in Fig.~\ref{fig:planck_error_T0} using a Fisher-matrix analysis (see Appendix~\ref{sec:method} for more details), the incorrect calibration method leads to a much larger uncertainty on $A_{\rm s}$, due to the $T_0$--$A_{\rm s}$ degeneracy, and a significantly smaller uncertainty on $T_0$, due to the artificial effect on lensing. Indeed, the incorrect-calibration error on $A_{\rm s}$ is dominated by that of $T_0$ at large $\ell_{\rm max}$, as expected 
given the $T_0$--$A_{\rm s}$ degeneracy.

This issue with the temperature calibration of theoretical CMB power spectra was a problem~\cite{GY_email,JL_email} in Refs.~\cite{T0censorship} and~\cite{2018CORE}, and affected the results in the latter reference since it found that the uncertainty on $A_{\rm s}$ increased substantially when including $T_0$ as a parameter.  It was also a problem in version~1 of Ref.~\cite{T0H0}, which led to an unrealistically tight predicted constraint on $T_0$ from CMB anisotropies.

In summary, since the CMB power spectra are most naturally dimensionless in theory and are also measured directly in $\Delta T/T$ units, there is no compelling need to present the spectra in temperature units. The current convention is simply one of convenience. For clarity of the physical meaning of the quantities being calculated and measured and for the sake of uniformity across power spectra, it would be better to present all future results as dimensionless quantities. In the rest of this paper, for all our analyses, we will follow Eq.~(\ref{eq:C_l unit}), the second approach, whenever we need to calculate $C_{\ell}$ in temperature units to meet the current convention.

\section{\texorpdfstring{\boldmath{$T_0$}}{T0} in cosmological models}\label{T0inCosmo}

With the clarification of temperature calibration in Sec.~\ref{sec:calibration}, it is clear that when we discuss changing $T_0$ in a 7-parameter $\Lambda$CDM model, we should only consider its impact on the dimensionless CMB power spectra $C_{\ell}^{\rm [D]}$ in \eq({\ref{eq:C_l unit}}). The investigation of the effects of varying $T_0$ on the CMB power spectra goes back at least to the paper by Hu et al.\ in 1995~\cite{HSSW} and was further developed about 12 years later~\cite{Chluba2008,Hamann2008}, with the constraining power of CMB power spectra on $T_0$ first explored using \textit{Planck} anisotropy data in section~6.7.3 of Ref.~\cite{2015Planckparameter}. However, it is only through the recent papers by Ivanov et al.~\cite{T0H0} and Bose \& Lombriser~\cite{openhotteruniverse} that the role of $T_0$ as a cosmological parameter has been more fully explored.

\subsection{The role of \texorpdfstring{\boldmath{$T_{\gamma}$}}{T0}} \label{role_of_Tgamma}

The temperature of the background photons, $T_{\gamma}$, is a function of time or redshift, with $T_{\gamma}(z=0)=T_0$. One can consider several different quantities to trace the evolution of the cosmological model, including scale factor $a$, redshift $z$, cosmological (or proper) time $t$, and Hubble parameter $H$. As a dynamical variable, $T_{\gamma}$ plays a similar role, and can be used as an alternative quantity to set the timescale. One can imagine different observers living in a particular Friedmann universe but observing the CMB at different times~\cite{ZMS2007,MZS2008}, with the observed $T_0$ as one way of fixing the epoch.

$T_{\gamma}$ also traces the overall radiation content of the Universe, which includes other light species that evolve in the same way as photons, often parameterised with $N_{\rm eff}$. Important epochs, such as nucleosynthesis or recombination, are set by a comparison between fundamental physical quantities (e.g., particle number densities, masses, and couplings) and the radiation content in the background cosmological model, which is set by $T_\gamma$. This ``clock'' can be made manifestly dimensionless (see Ref.~\cite{Narimani2012}) by defining the quantity $\Theta=kT_{\gamma}/m_{\rm p}c^2$, the ratio of thermal energy to proton mass.

In the $\lc$ model, the expansion history of the Universe is determined by the energy densities of baryons $\rho_{\rm b}$, cold dark matter $\rho_{\rm c}$, radiation $\rho_{\rm r}$, and dark energy $\rho_{\Lambda}$, most of which vary with time. To relate the densities at a specific epoch with the present-day cosmological parameters $\Omega_{\rm b} h^2$, $\Omega_{\rm c}h^2$, and $T_0$, we have, for example,
\begin{equation}
\rho_{\rm b}= \rho_{b,0}a^{-3}=\rho_{b,0}(T_0/T_{\gamma})^{-3}\propto \frac{\Omega_{\rm b} h^2}{(T_0/\overline{T}_{0, \rm F})^3}T_{\gamma}^3.
\label{eq:bbequation}
\end{equation}
We can set $\widetilde{\omega}_{\rm b}=\Omega_{\rm b} h^2(T_0/\overline{T}_{0, \rm F})^{-3}$, where $\overline{T}_{0, \rm F}$ (the central value of the FIRAS measurement; see Appendix~\ref{sec:tempdefs}) is used to allow easy comparison between the 6- and 7-parameter $\lc$ models, such that when $T_0=\overline{T}_{0, \rm F}$, then $\widetilde{\omega}_{\rm b}=\Omega_{\rm b} h^2$. According to \eq({\ref{eq:bbequation}}), $\rho_{\rm b}\propto \widetilde{\omega}_{\rm b} T_{\gamma}^3$, and, similarly, we can define $\widetilde{\omega}_{\rm c}=\Omega_{\rm c} h^2(T_0/\overline{T}_{0, \rm F})^{-3}$, so that $\rho_{\rm c}\propto \widetilde{\omega}_{\rm c} T_{\gamma}^3$. The radiation density $\rho_{r}$ is sufficiently described by $T_{\gamma}$ and $N_{\rm eff}$ (we take $N_{\rm eff}\,{=}\,3.046$ here). Therefore, when we use $T_{\gamma}$ as the timescale, the expansion history $H$ of the early Universe can be written as a function of $\bb$, $\bc$, and $T_{\gamma}$ only, with dark energy negligible at early times.

The redefined present-day cosmological parameters $\widetilde{\omega}_{\rm b}$ and $\widetilde{\omega}_{\rm c}$ can be used as alternatives to $\bht$ and $\cht$ in a 7-parameter extension of $\lc$ with $T_0$ as a free variable. In fact, $\bb$ and $\bc$ in $\lc{+}T_0$ play a similar role to $\bht$ and $\cht$ in 6-parameter $\lc$. Using a Fisher-matrix analysis with $TT$, $TE$, and $EE$ power spectra (as described in Appendix~\ref{sec:method}), we can see in Fig.~\ref{fig:bbbcequivalence} that with negligible instrumental noise and perfect removal of foreground signals (i.e., the CVL assumption) the constraints we obtain for $\bb$ are essentially the same in the two models.  Although still quite similar, there is some noticeable difference for $\bc$, due to the effects of lensing. As seen in Fig.~\ref{fig:bbbcequivalence}, the $\bc$ constraints for the two models begin to differ at $\ell_{\rm max}\simeq 1000$ and then start to converge again at $\ell_{\rm max}\simeq 4000$. This behaviour can be explained by the transition between different lensing effects: lensing smoothing, which dominates out to $\ell$ of a few thousand; and extra small-scale power, which dominates the spectra at higher $\ell$, as diffusion damping wipes out the primary power spectra~\cite{06LewisLensing}.  The parameter $\bc$ exhibits much stronger degeneracy with $T_0$ in the lensing-smoothing regime than does $\bb$, thereby weakening the constraining power of $\bc$ in $\lc$+$T_0$.  This degeneracy is removed in the small-scale lensing power regime, providing additional constraining power for $\bc$, independent of $T_0$.  While the uncertainty on $\bc$ is clearly impacted by opening up $T_0$ as a free parameter, since the difference between the $\bc$ constraints of the two models is relatively small in such a CVL setting, we can for the most part regard $\bc$ as a replacement for $\cht$ in a 7-parameter $\lc$+$T_0$ model.  For the other four standard cosmological parameters, namely $\{\theta_{\ast}$, $A_{\rm s}$, $n_{\rm s}$, $\tau\}$, the forecast errors for each parameter only increase slightly by changing $T_0$ from a constant to a free parameter, under the CVL assumption.

\begin{figure}[htbp!]
\centerline{\includegraphics[width=\hsize]{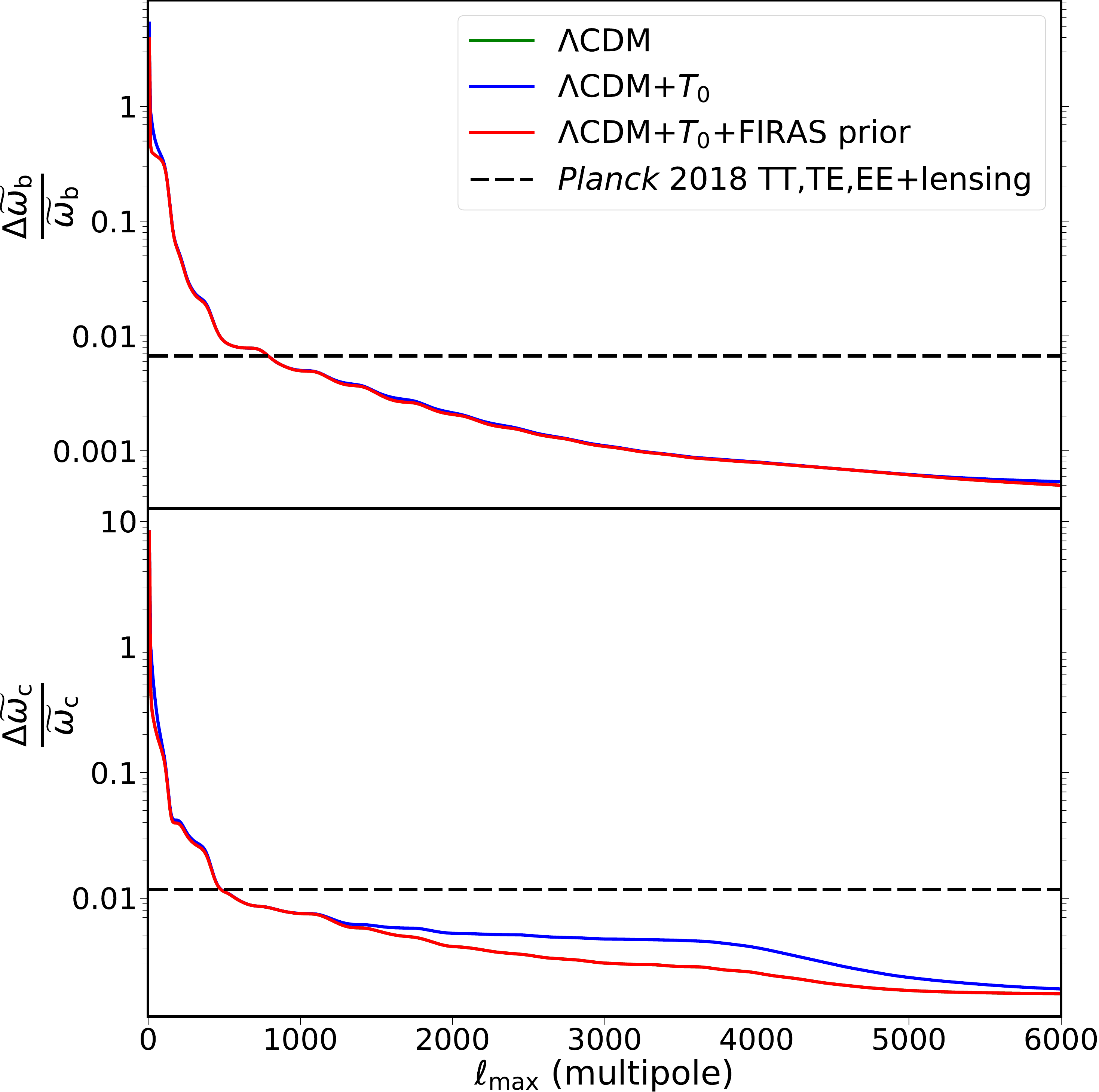}}
\caption{Forecast constraints on $\bb$ and $\bc$ as a function of the maximum multipole considered.  We specifically give the
forecasts with negligible instrumental noise (i.e., the cosmic-variance limit, described in Appendix~\ref{sec:method}) from a combination of CMB $TT$, $TE$, and $EE$ power spectra for several different assumptions. The blue curve shows the constraints for a 7-parameter $\lc$ model with $T_0$ as a free variable. The red curve shows the constraints for a 7-parameter $\lc$ model, with $T_0$ having a FIRAS prior (i.e., following \eq\ref{eq: FIRAS_Error}). The green curve shows the constraints for the standard 6-parameter $\lc$ model, where $\bb=\bht$ and $\bc=\cht$.  The green curve completely overlaps with the red curve, which means that when applying the FIRAS prior, there is no discernible difference for $\bb$ and $\bc$ between the 6- and 7-parameter models. The blue curve basically follows the red curve for $\bb$, while there is some noticeable difference between the two curves for $\bc$. This difference can be explained by the impacts of lensing smoothing and small-scale lensing power (see text).}
\label{fig:bbbcequivalence}
\end{figure}

\subsection{Recombination physics and \texorpdfstring{\boldmath{$T_0$}}{T0}} \label{role_of_T0}

During the major events of the early Universe (such as nucleosynthesis and recombination), the physical processes are largely determined by the number densities of electrons and baryons, as well as the energy density of cold dark matter, together with the expansion history (which at very early times is predominantly driven by the radiation component). With the known average mass of a baryon, we can use $\rho_{\rm b}$ as a proxy for the associated number density, so that $\rho_{\rm b}\propto \widetilde{\omega}_{\rm b} T_{\gamma}^3$, $\rho_{\rm c}\propto \widetilde{\omega}_{\rm c} T_{\gamma}^3$, and $T_{\gamma}$ are the only relevant background quantities in the $\lc$ model determining the progress of decoupling of particle interactions. From such a qualitative argument, it is already clear that all the physical processes in the early Universe are set by $\bb$ and $\bc$, without explicit dependence on $T_0$.  Indeed this notion that recombination depends only on physics local to last scattering was the basis of the use of effective models, with varying $T_0$, to calculate CMB spectra in models with large underdensities in Ref.~\cite{mzs10}.  Nonetheless, we will now offer a more detailed discussion of the process of recombination and the epoch of last scattering (since they are of utmost importance to CMB power spectra) as examples to further illustrate the idea. 

Cosmological recombination is described by a set of coupled first-order differential equations for the free electron fraction $X_{\rm e}$, the electron temperature $T_{\rm e}$,\footnote{$T_{\rm e}$ is very close to $T_\gamma$ at early times, differing only at the level of $10^{-7}$ during recombination~\cite{ScottMoss}.} and the photon phase-space density, with respect to proper time $t$~\cite{P1968,ZKS,CosmoRec,Hyrec}. Using the differential relation
\begin{equation}
    dt=\frac{dt}{da}\frac{da}{dT_{\gamma}}dT_{\gamma}=-\frac{1}{HT_{\gamma}}dT_{\gamma},
    \label{t-Trelation}
\end{equation}
which does not explicitly depend on $T_0$, we can rewrite the governing differential equations for recombination in terms of $T_{\gamma}$~\cite{T0H0}; it is then clear that the recombination history $X_{\rm e}(t)$ is determined only by the parameters $\bb$, $\bc$, and $T_{\gamma}$, along with physical constants.

The differential Thomson optical depth for recombination is given by \begin{equation}
\dot{\tau}_{\rm rec} = \frac{d\tau_{\rm rec}}{dt} = c\sigma_{\rm{T}} n_{\rm{H}}(T_{\gamma},\bb) X_{\rm e}(T_{\gamma},\bb,\bc),
\end{equation}
where $c$ is the speed of light, $\sigma_{\rm{T}}$ is the Thomson cross-section, and $n_{\rm{H}}$ is the number density of hydrogen atoms, which can be converted to $\rho_{\rm{b}}$ with a fixed helium fraction.  We use the subscript ``rec'' to distinguish the optical depth coming from the recombination process from the small optical depth coming from the reionization parameter $\tau$ in $\lc$.  The optical depth for recombination out to some specific epoch is then simply
\begin{align}
    \tau_{\rm rec}&=\int_{t}^{t_0}\dot{\tau}_{\rm rec}(t') dt'=\int_{T_0}^{T_{\gamma}}\frac{1}{T_{\gamma}'H}\dot{\tau}_{\rm rec}(T_{\gamma}') dT_{\gamma}'\nonumber\\
    &=c\sigma_{\rm{T}}\int_{T_0}^{T_{\gamma}}\frac{n_{\rm{H}}(T_{\gamma}',\bb) X_{\rm e}(T_{\gamma}',\bb,\bc)}{T_{\gamma}'H(T_{\gamma}',\bb,\bc)} dT_{\gamma}',
\label{eq:recom_optical_depth}
\end{align}
where $t_0$ is the current proper time.

There are two common approaches for defining the epoch of last scattering $t_{\ast}$ (or similarly $z_{\ast}$ or $T_{\ast}$). The first approach is to define it to correspond to the peak of the visibility function $g(t)=\dot{\tau}_{\rm rec}\exp(-\tau_{\rm rec})$ (e.g., Ref.~\cite{WMAP1YearCosmologicalParameter}). The peak condition $dg(t_{\ast})/dt=0$ yields
\begin{equation}
\frac{d\dot{\tau}_{\rm rec}(t_{\ast})}{dt}=\dot{\tau}_{\rm rec}^2(t_\ast).
\label{eq:last_scattering_visbility}
\end{equation}
Recasting this in terms of $T_{\gamma}$, we obtain a precise definition of the last-scattering temperature $T_{\ast}$. Since both $dt/dT_{\gamma}$ and $\dot{\tau}_{\rm rec}$ are independent of $T_0$, then $T_{\ast}$ is only a function of $\bb$ and $\bc$, without any explicit $T_0$ dependence.

The second approach is to define the last-scattering epoch to correspond to the time back to when the optical depth reaches unity, i.e., $\tau_{\rm rec}(t_{\ast})=\tau_{\rm rec}(T_{\ast})=1$ (e.g., Refs.~\cite{WH2005,CAMB,2018Planckparameter}). In this case, at least in principle, $\tau_{\rm rec}$ does depend on $T_0$ through the lower bound of the integration in Eq.~(\ref{eq:recom_optical_depth}); however, this dependence is extremely weak, since $X_{\rm e}$ (excluding the effect of reionization) is only of order $10^{-4}$ at the current epoch, thereby giving negligible contribution to $\tau_{\rm rec}$ during the late-time evolution of the Universe~\cite{ZMS2007}.  We numerically assessed the degree of this weak $T_0$ dependence, finding that with either a fixed integrand in Eq.~(\ref{eq:recom_optical_depth}) or a fixed $\theta_{\ast}$ value (detailed definition of the parameter $\theta_{\ast}$ is given in Sec.~\ref{othercosmoparameters}),\footnote{We briefly compare the two methods used to assess the impact of $T_0$ on $T_{\ast}$ in Eq.~(\ref{eq:recom_optical_depth}). Since $T_0$ only appears as the lower bound of the integration, the obvious method is to fix the integrand while varying $T_0$, which corresponds with changing the current observation epoch in the same Friedmann model. However, as discussed in Sec.~\ref{othercosmoparameters}, it is more physical to fix $\theta_{\ast}$ when we analyse the impact of $T_0$ under CMB power spectra constraints, so we actually performed the check using both methods.} a $15\,\%$ change of $T_0$ around $\overline{T}_{\rm 0,F}$ leads to no discernible change in $T_{\ast}$. For all practical purposes, we can therefore consider $T_{\ast}$ to be independent of $T_0$ in Eq.~(\ref{eq:recom_optical_depth}).  In practice, both these definitions of the last-scattering epoch yield almost the same results, with the calculated results for $T_{\ast}$ (or $z_{\ast}$) under the two approaches coinciding within $5\times10^{-4}$ in a fiducial model near the current measurement of cosmological parameters~\cite{WH2005}.  We adopt the second approach in this paper for definition of the last-scattering epoch to allow for later comparison with $\textit{Planck}$ results.

Using a Markov-chain Monte Carlo (MCMC) method with the 2018 {\em Planck} TT,TE,EE+lensing likelihood in a 7-parameter $\lc{+}T_0$ model, we find that $T_{\ast}=2970.9^{+0.8}_{-1.0}\,$K, which is constrained extremely well, to 0.03\,\%.  This is consistent with the $\lc$ $T_*$ value, as expected based on the above physical arguments that $T_*$ depends only on conditions local to last scattering; however, it contradicts the claim of a changing temperature at last scattering in the $\lc{+}T_0$ model made in Ref.~\cite{20RephaeliT0}.  Coincidentally the uncertainty of $T_\ast$ is comparable to that of the FIRAS result on $T_0$ in Eq.~(\ref{eq: FIRAS_Error}).   In contrast to this, we find $z_{\ast}=936\pm 60$, with a relatively large uncertainty. Such great uncertainty is mainly due to the weak constraint on $T_0$ from CMB power spectra, which will be further discussed in Sec.~\ref{constraining T0}. In the standard 6-parameter $\lc$ model, the relative uncertainty on $z_\ast$ is the same as that on $T_\ast$.  Nevertheless, we can see that, when allowing $T_0$ to be a free parameter, the physics more directly constrains the value of $T_\ast$ than $z_\ast$ from recombination. As established in Fig.~\ref{fig:bbbcequivalence}, $\widetilde{\omega}_{\rm b}$ and $\widetilde{\omega}_{\rm c}$ are constrained by CMB power spectra to almost the same accuracy in a 7-parameter $\lc{+}T_0$ model as $\bht$ and $\cht$ are in 6-parameter $\lc$. With the current \textit{Planck} likelihood, $\bb$ and $\bc$ are constrained to the percent level (see the dashed lines in Fig.~\ref{fig:bbbcequivalence}), in contrast to the 0.03\,\% accuracy for $T_{\ast}$. Compared to the baryon and cold dark matter densities, the temperature at last scattering is much more tightly constrained, since the last-scattering epoch $T_{\ast}$ (or $z_{\ast}$ in $\lc$) is generally a weak function of cosmological parameters~\cite{HuSmall-Scale1996}.

With $T_{\ast}$ as an extremely well measured quantity, then the baryon and cold dark matter densities at last scattering, $\rho_{\rm b}(T_\ast)\propto \bb T_\ast^3$ and $\rho_{\rm c}(T_\ast)\propto \bc T_\ast^3$, are in fact relatively well constrained by CMB power spectra. Some of the papers in the literature discussing $T_0$ as a free parameter seem to suggest that $T_0$ directly impacts recombination~\cite{Chluba2008,2018CORE,20RephaeliT0}. This is mainly due to directly varying $T_0$ without considering $\bb$ and $\bc$ as new variables. If one simply fixes $\bht$ and $\cht$, while changing $T_0$, the baryon and matter densities at the recombination epoch are actually changed, which drastically alters the recombination history~\cite{Chluba2008}.  However, baryon and dark matter densities at last scattering are actually fairly well measured through recombination physics, so such an approach of fixing $\bht$ and $\cht$ and changing $T_0$ does not correspond with the constraint conditions set by CMB power spectra.  A better and more physically motivated approach for analysing the effects of $T_0$ on cosmology is to fix $\bb$ and $\bc$, which are well constrained in the 7-parameter $\lc{+}T_0$ model.

When analyzing the impact of varying $T_0$ in cosmological models, it is crucial to distinguish $T_0$ from $T_\ast$, based on their roles in recombination. The dynamical variable $T_{\gamma}$ clearly enters recombination physics, with $T_\ast$ being set by the absolute energy scale of atomic and particle physics and well determined by CMB power spectra; however, the present-day background photon temperature $T_0$ only serves as a relative timescale that indicates how much the Universe has expanded since the time of last scattering, with $T_0$ itself having no effect on the physics of the early Universe when using the physically relevant parameters $\bb$ and $\bc$.

\subsection{Constraining \texorpdfstring{\boldmath{$T_0$}}{T0} with CMB power spectra}\label{constraining T0}
As established in Sec.~\ref{sec:calibration}, the calibration of CMB anisotropy data does not actually involve the current experimentally determined CMB monopole, $T_0$.  Nevertheless, if $T_0$ impacts the dimensionless $C_{\ell}$, then in principle we can provide an independent constraint on $T_0$ from the measured (dimensionless) anisotropy power spectra.

Since the {\em shapes} of the primary CMB anisotropy power spectra are determined by the physical conditions around the last-scattering epoch (on which $T_0$ has no impact when  $\bb$ and $\bc$ are used as density parameters), they only tell us about the physical conditions at recombination and provide no constraints on $T_0$. Nevertheless, changing $T_0$ does alter the amount of expansion after the last-scattering epoch, changing the angular-diameter distance $D_{\rm A}$ back to the recombination epoch, which shifts the anisotropies in multipole space.  However, the angular scale of the acoustic oscillations is actually extremely tightly constrained by CMB experiments~\cite{2018Planckparameter}. Therefore, as we vary $T_0$, we have to take advantage of the geometrical degeneracy and change the values of other cosmological parameters in order to preserve the angular acoustic scale (as discussed further in the next section). As a result, the acoustic oscillations in the power spectra, both in shape and period, give no information about the epoch at which we are observing.

By contrast, secondary anisotropies, generated after the last-scattering epoch, can break this degeneracy and provide some constraints on $T_0$. In particular as structure forms and dark energy starts to dominate at low redshifts, both the so-called ``integrated Sachs-Wolfe'' (ISW) effect~\cite{SW1967} and gravitational lensing depend on the late-time expansion and give observable imprints on the CMB anisotropies, and hence their signatures are able to constrain $T_0$.  To quantify these effects, we calculate the predicted $T_0$ constraints using lensed $TT$, $TE$, and $EE$ power spectra through a Fisher-matrix analysis (see Appendix~\ref{sec:method}). We can see in Fig.~\ref{fig:planck_error_T0} that as $\ell_{\rm max}$ is increased, there is a substantial drop in the uncertainty on $T_0$ until $\ell_{\rm max}\simeq30$ (which is where the ISW effect contributes) and a continuous decrease towards higher multipoles (where gravitational lensing contributes).

\begin{figure}[htbp!]
\centerline{\includegraphics[width=\hsize]{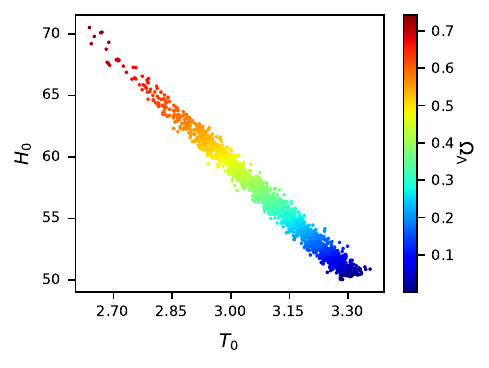}}
\caption{Samples from MCMC posteriors using the 2018 {\em Planck} TT,TE,EE+lensing likelihood for the 7-parameter $\lc{+}T_0$ model, plotted in the $T_0$--$H_0$ plane and colour coded by the value of $\Omega_\Lambda$. The degeneracy between $T_0$ and $H_0$ is clearly seen in the plot, with $\Omega_\Lambda$ corresponding to different amounts of late-time expansion and late-ISW effects. The current constraint on $T_0$ from the $\textit{Planck}$ data set is highly dependent on the prior choice, with $\Omega_\Lambda\geq0$ imposed here.}
\label{fig:T0-H0}
\end{figure}

Using the 2018 {\em Planck} TT,TE,EE+lensing likelihood, we find the $68\,\%$ interval to be $T_0=3.10^{+0.18}_{-0.09}\,$K and the $95\,\%$ interval to be $T_0=3.11^{+0.22}_{-0.25}\,$K, with the distribution of $T_0$ values shown in Fig.~\ref{fig:T0-H0}. Our results were found using an MCMC analysis with the \texttt{Cobaya}~\cite{2020Cobaya} and \texttt{CAMB}~\cite{CAMB} codes, with \texttt{Recfast}~\cite{1999Recfast,2000Recfast} included for the recombination modelling. These results are in excellent agreement with those in Ref.~\cite{T0H0}, which used a different set of cosmological codes.  As seen in Fig.~\ref{fig:T0-H0}, this constraint depends strongly on the priors imposed on the underlying cosmology, most prominently the assumption that the dark energy should be positive, i.e., $\Omega_\Lambda\geq 0$. For models with negative dark energy $T_0$ can go as high as $3.8\,$K, and the associated uncertainty on $T_0$ becomes significantly larger, with a $1\,\sigma$ value around $0.2\,$K, which agrees with the blue line in Fig.~\ref{fig:planck_error_T0} determined by the Fisher forecast. In general, the current \textit{Planck} data provide a poor constraint on $T_0$. Since $\bb$ and $\bc$ are as well constrained in a 7-parameter $\lc{+}T_0$ model as $\bht$ and $\cht$ are in the 6-parameter $\lc$ model, the weak constraint on $T_0$ leads to considerable degeneracy between $T_0$ and $\bht$ and between $T_0$ and $\cht$~\cite{2015Planckparameter}.

As seen above, the central value of the $T_0$ constraint from the \textit{Planck} 2018 data is about 2$\,\sigma$ higher than the FIRAS measurement. This slight shift is mainly caused by the twin facts that the low multipoles ($\ell \lesssim 30$) in the measured $TT$ power spectrum of $\textit{Planck}$ are somewhat low compared to the best-fit $\Lambda$CDM model and that the observed $TT$ power spectrum appears to be more smoothed by lensing than expected (as usually characterized by the consistency parameter $A_{\rm L}$)~\cite{2018Planckparameter}. Both the ``low-$\ell$ deficit'' and ``$A_{\rm L}$ tension'' have already been shown to have an impact on the standard parameter constraints~\cite{Planckshifts,2018Planckparameter}. With a higher $T_0$ value, the theoretical $TT$ power spectra prefer less large-scale power and more lensing smoothing~\cite{T0H0}, which both correspond with the direction of the deviation in the observed data. Since $T_0$ is only constrained by late-ISW and lensing effects, the low-$\ell$ deficit and $A_{\rm L}$ tension lead to a small shift in the parameter constraint on $T_0$, resulting in this 2$\,\sigma$ deviation from the FIRAS result.

\begin{figure}[htbp!]
\centerline{\includegraphics[width=\hsize]{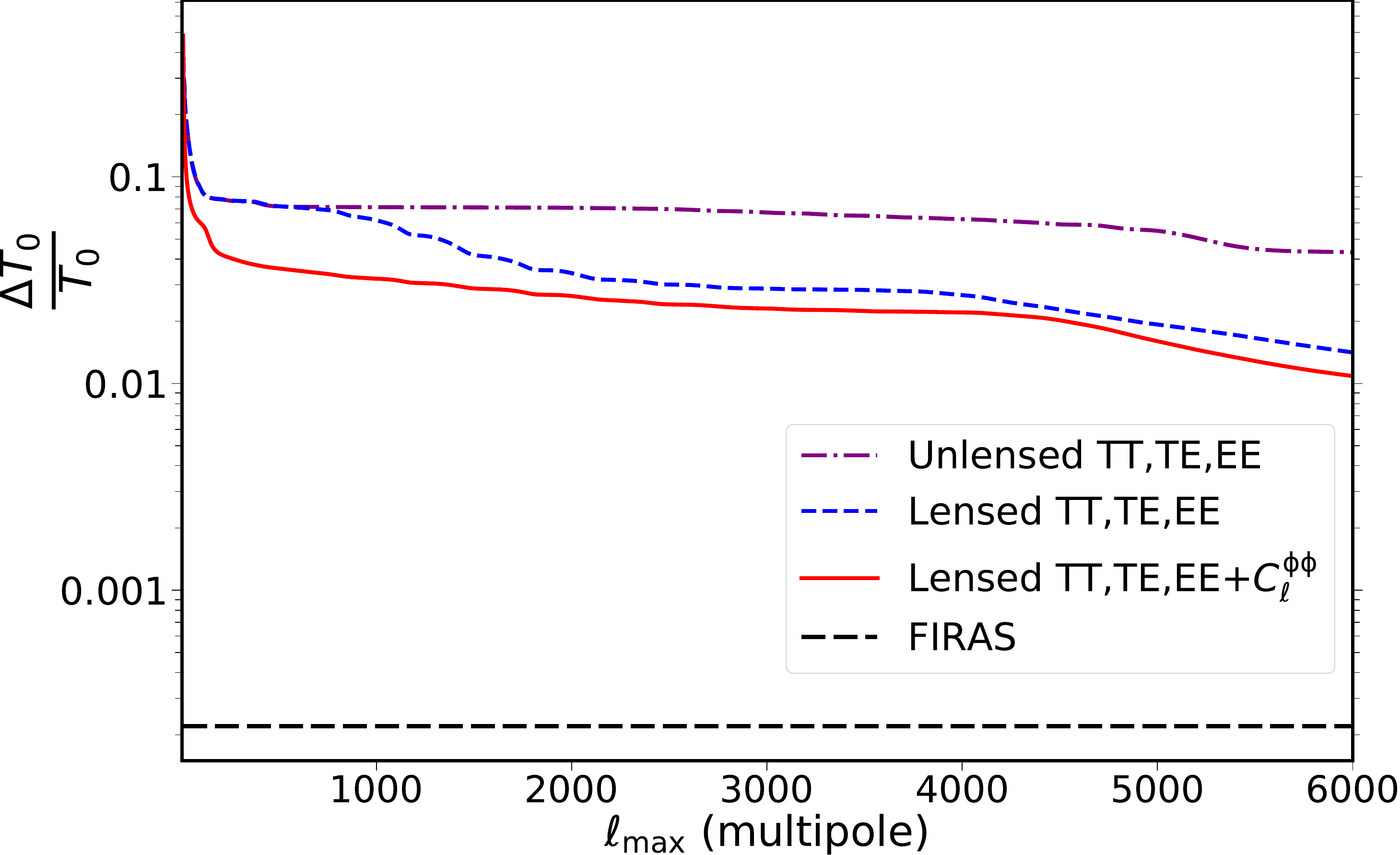}}
\caption{Predictions of the level of uncertainty on $T_0$ as a function of maximum multipole, coming from CMB anisotropy power spectra, under the cosmic-variance limit (described in Appendix~\ref{sec:method}). The purple (dot-dashed) curve corresponds to the use of unlensed $TT$, $TE$, and $EE$ power spectra (although unrealistic, they are plotted here to assess the size of the different physical effects), the blue (dashed) curve is for the lensed $TT$, $TE$, and $EE$ power spectra, and the red (solid) curve uses the lensed $TT$, $TE$, and $EE$ power spectra as well as the lensing-reconstruction power spectrum $C_{\ell}^{\phi\phi}$. For comparison, the black line shows the uncertainty of the FIRAS measurement.}
\label{fig:lensing_T0}
\end{figure}
 
Figure~\ref{fig:lensing_T0} demonstrates the predicted constraining power on $T_0$ coming from CMB power spectra for an ideal CMB experiment (described in Appendix~\ref{sec:method}). With negligible instrumental noise, gravitational lensing provides a much stronger constraint on the amount of late-time Universe expansion compared to the constraint coming from \textit{Planck}; this is because the temperature and polarization power spectra for $\ell\gtrsim1000$ (where lensing smoothing plays a major role) and the lensing reconstruction spectrum $C_{\ell}^{\phi\phi}$ are much better measured in an ideal experiment, while the late-ISW effect is limited by the high cosmic variance at low $\ell$. In general, lensing decreases the uncertainty on $T_0$ (out to $\ell_{\rm max} \simeq 6000$) by around a factor of 4 compared with the late-ISW effect.  The lensing smoothing effect starts to become important at $\ell\gtrsim1000$ as expected, and additionally the lensing reconstruction spectrum drives down the $T_0$ uncertainty mainly around $\ell\simeq100$, where the peak of $C_{\ell}^{\phi\phi}$ is located. Changes in cosmological parameters mainly lead to a shift of the overall amplitude for $C_{\ell}^{\phi\phi}$ (rather than a change in the shape~\cite{Smith2006}), so the peak region provides most of the constraining power for any parameter. Combining both lensing smoothing and lensing reconstruction leads to an improvement in the the $T_0$ constraint by a small amount. However, even the most ideal CMB experiment is only able to constrain $T_0$ to around $0.03\,$K, which is still about two orders of magnitude poorer than the FIRAS measurement. This means that we will never be able to obtain a competitive constraint on $T_0$ from CMB anisotropy data alone, compared to FIRAS.

It {\it is\/} possible to obtain a better constraint on $T_0$ by combining CMB anisotropy data with other cosmological data sets. Reference~\cite{2015Planckparameter} has already noted that combining baryonic acoustic oscillation (BAO) data~\cite{2017BAODR12} with the full 2015 \textit{Planck} likelihood leads to a $68\,\%$ confidence interval of $T_0=(2.718 \pm 0.021)\,{\rm K}$, and we find a similar result when combining the \textit{Planck} 2018 likelihood with current BAO data. Unlike the constraint from \textit{Planck} data alone, this measurement is not shifted high, but is completely compatible with the FIRAS measurement. Because $T_0$ is effectively a parameter measuring the amount of expansion since recombination, any measurement of the late-time Universe that constrains the expansion history should strengthen the constraint on $T_0$. With the error from the noise-free CMB data alone also predicted to be at around the $1\,\%$ level, one might expect to obtain an independent $T_0$ measurement with a precision comparable to the current FIRAS uncertainty level by combining future CMB anisotropy experiments with upcoming large-scale structure (LSS) surveys such as DESI~\cite{2016DESIintro} and \textit{Euclid}~\cite{19EuclidForecast}. 

To give a simplistic assessment of this idea, we combine the Fisher forecast from $TT$, $TE$, $EE$, and lensing reconstruction spectra in an ideal CMB anisotropy experiment (shown in Fig.~\ref{fig:lensing_T0}) with the forecast from measurements of the BAO scale in a \textit{Euclid}-like survey~\cite{Euclid2011}. Details of our BAO forecast method are given in Appendix~\ref{sec:method}. The combination of CMB and BAO data is predicted to constrain $T_0$ with an uncertainty of around $0.006\,$K, which is about 4 times better than the constraint from ideal CMB alone, or the constraint from a combination of \textit{Planck} and current BAO data. Even though $0.006\,$K is an order of magnitude less constraining than the FIRAS measurement, the CMB anisotropy plus BAO constraint will nonetheless provide a relatively precise measurement of $T_0$ that is independent of FIRAS. In principle, we could further drive down the uncertainty on $T_0$ with the full constraining power of a $\textit{Euclid}$-like experiment by including galaxy clustering and weak lensing measurements, in addition to BAO data.  However, determining such future predictions from LSS data is outside the scope of this paper. Still, from our discussion of the role of $T_0$ and our simple combination of future BAO and ideal CMB anisotropy constraints, it is already clear that combining next-generation CMB and LSS data will offer an independent measurement of $T_0$ with a precision that could approach that of FIRAS.  Whether such a constraint will ever surpass the current FIRAS error remains to be seen.

\subsection{\texorpdfstring{\boldmath{$T_0$}}{T0} and other cosmological parameters}\label{othercosmoparameters}

Since CMB data alone only provide a weak constraint on $T_0$, through the late-ISW and lensing effects, it is possible to connect $T_0$ variation with other cosmological parameters.  For example Ivanov et al.~\cite{T0H0} focus on the tension between CMB-derived~\cite{2018Planckparameter} and distance-ladder-derived estimates of the Hubble constant~\cite{2019SH0ES}.  Although they conclude that changing $T_0$ does not provide a viable solution to this apparent cosmological parameter discrepancy, it is nevertheless instructive to see how the parameter degeneracies work here.

In the standard $\lc$ model, $\theta_{\ast}$, which is the ratio between the proper sound horizon at recombination ($d^\ast_{\rm s}=a_*r_s$) and the angular-diameter distance back to the recombination epoch ($D_{\rm A}$), is the best constrained parameter from CMB experiments due to the well-measured acoustic oscillations of the CMB power spectra~\cite{2018Planckparameter}. The angular-diameter distance to the last-scattering epoch can be written as
\begin{align}
D_{\rm A}&=\frac{1}{(1+z_\ast)}\int_{0}^{z_\ast}\frac{c\,dz}{H(z,\Omega_{\rm m},\Omega_\Lambda)}\nonumber\\
&=\frac{1}{T_\ast}\int_{T_0}^{T_\ast}\frac{c\,dT_{\gamma}}{H(T_{\gamma},H_0,\bb,\bc)},
\label{DAequation}
\end{align}
where we neglect the radiation, neutrino, and curvature background components for simplicity, since the effects of these components on the expansion history are relatively small compared to the impacts of varying $T_0$ within the weak $T_0$ constraint from \textit{Planck}. Reference~\cite{T0H0} demonstrates that $d^\ast_{\rm s} = (T_0/T_\ast)r_{\rm s}$ is only a function of $\bb$ and $\bc$, without explicit dependence on $T_0$. Therefore, since $\bb$, $\bc$, and $\theta_{\ast}$ are measured significantly better through CMB power spectra compared to $T_0$, we can regard $\bb$, $\bc$, and $\theta_{\ast}$ as fixed here, so that $T_\ast$, $d^\ast_{\rm s}$, and $D_{\rm A}$ can also be considered to be fixed. Viewed as an integral with respect to $T_{\gamma}$, $D_{\rm A}$ can now be treated as a function of only two variables (namely $T_0$ and $H_0$) whose combination in some functional form  gives a fixed value. Hence, $T_0$ and $H_0$ have an approximate degeneracy when $\bb$, $\bc$, and $\theta_{\ast}$ are measured so well compared to the poor constraint on $T_0$ coming from CMB power spectra. This $T_0$--$H_0$ degeneracy can be best seen in Fig.~\ref{fig:T0-H0} with samples from an MCMC run. The uncertainty on $T_0$ physically corresponds with the uncertainty on the amount of dark energy contained in the Universe, which is demonstrated by the variation of $\Omega_\Lambda$ along the $T_0$--$H_0$ degeneracy line in the plot. Since $\theta_{\ast}$ is well measured by CMB experiments, we need to change the amount of dark energy in the Friedmann model to compensate for the change of $T_0$, in order to obtain a fixed $D_{\rm A}$ value. 

It might be tempting to relate this $T_0$--$H_0$ degeneracy to the Hubble-constant tension. Reference~\cite{T0H0} finds $T_0=(2.56\pm0.05)\,{\rm K}$ by combining \textit{Planck} 2018 data with SH0ES, where the $H_0$ measurement from the distance ladder provides an additional constraint on the late-time expansion. However, such a result deviates from the current best measurement $T_{0,{\rm F}}=(2.7255\pm0.0006)\,{\rm K}$ by hundreds of $\sigma$ in terms of the FIRAS uncertainty. Therefore, addressing the Hubble tension by varying $T_0$ requires us to completely discount the FIRAS measurement. One should remember that FIRAS measured the entire frequency spectrum of the CMB, rather than just a single value of the absolute intensity, with the shape of this spectrum determining the value of $T_0$; hence it is hard to imagine how the derived value of $T_0$ could be so far off. Moreover, the experimental determination of $T_0$ does not come entirely from FIRAS, since there have been many other measurements of the CMB monopole temperature.   Although the current $0.02\,\%$ uncertainty in $T_0$ is dominated by data from FIRAS, there are other independent measurements with uncertainties that are still impressively small~\cite{GHW1990,SB2002}. Reference~\cite{FIRAStemp} provides an alternative estimate from a compilation of measurements excluding FIRAS, $T_0 = (2.729 \pm  0.004)\,{\rm K}$, which still has approximately 0.1\,\% precision (and is in excellent agreement with the FIRAS constraint).  Therefore, the reasonable range of variation of $T_0$ that is consistent with experimental constraints has an entirely negligible effect on $H_0$, with or without the {\it COBE}-FIRAS data.  Indeed, since both are background variables, we would expect to require a temperature shift of order the Hubble parameter discrepancy, $\Delta T_0/T_0 \sim \Delta H_0/H_0 \sim 10\,\%$, for the Hubble tension to be resolved with a shift in $T_0$.

On the other hand, since the constraint on $T_0$ coming from the CMB power spectra arises because of the late-ISW and lensing effects, there {\it will be\/} degeneracies between $T_0$ and other cosmological parameters in some extensions of standard $\lc$.  With \textit{Planck} 2018 data, allowing $T_0$ to be a free parameter can slightly relieve the so-called ``$A_{\rm L}$ tension''~\cite{openhotteruniverse}, since varying $T_0$ also changes the amount of lensing smoothing. Because the spatial curvature parameter $\Omega_K$, the total neutrino mass $\sum m_{\nu}$, and the dark energy equation of state parameter $w$ also contribute to the late-time expansion of the Universe (and therefore have a direct impact on the late-ISW effect and lensing~\cite{openhotteruniverse,neutrionolensing-Hall,T0censorship}), these parameters will also be somewhat correlated with $T_0$. 
For a counter-example, the parameter $N_{\rm eff}$ (the effective number of light particle species) only matters when neutrinos are relativistic in the early Universe and so will not be correlated with $T_0$.
Nevertheless, even in cases where there is some degeneracy, the actual uncertainty in the measured $T_0$ is so small that there is no way of using a shift in $T_0$ to resolve any apparent parameter tensions.

\section{Impact of FIRAS uncertainty}\label{ImpactFIRASerror}

\subsection{Current CMB anisotropy experiments}
\label{CurrentImpactFIRASerror}
Despite our pedagogical discussion on the role of $T_0$ as a free variable in Sec.~\ref{T0inCosmo}, any realistic treatment of temperature as a variable requires the adoption of the FIRAS measurement as a prior, representing our current knowledge of $T_0$. To fold in this information, we will adopt a Gaussian distribution with central value $\overline{T}_{0, \rm F} = 2.7255\,$K and standard deviation $\sigma_{\rm F}=0.0006\,$K as the prior on $T_0$. The conventional belief is that the FIRAS measurement does not make any appreciable difference to the current CMB results. This is true for the constraints on the base parameters of $\lc$, as well as most other parameters. However, as seen in Table~\ref{tab:CurrentFIRASError}, we noticeably underestimate the error for some of the derived parameters by ignoring the FIRAS uncertainty. Even for present-day anisotropy experiments, the FIRAS prior can have some unexpected impacts on our parameter constraints. 

\begin{table}[htbp!]
\setlength{\tabcolsep}{10pt}
\begin{tabular}{l c c}
\hline
\hline
\noalign{\vskip 1pt}
Parameter& $T_0$ fixed& FIRAS prior\\
\hline
\noalign{\vskip 2pt}
100$\theta_{\rm MC}$& $1.04092\pm0.00031$ & $1.04091\pm0.00037$\\
100$\theta_{\ast}$ & $1.04110\pm0.00031$ & $1.04111\pm0.00030$\\
$z_{\ast}$ & $1089.92\pm0.25$ & $1089.91\pm0.34$ \\
$T_{\ast}$ [K] & $2973.27\pm0.67$ & $2973.28\pm0.67$\\
$z_{\rm drag}$ &  $1059.94\pm0.30$ & $1059.94\pm0.37$\\
$T_{\rm drag}$ [K] & $2891.60\pm0.78$ & $2891.60\pm0.78$\\
\hline
\end{tabular}
\caption{68\,\% intervals for some derived parameters using the 2018 {\em Planck} TT,TE,EE+lensing likelihood in an MCMC analysis. We specifically look at $\theta_{\rm MC}$ (which we compare with $\theta_\ast$), $z_{\ast}$ and $T_\ast$ (the redshift and temperature at the last-scattering epoch), and $z_{\rm drag}$ and $T_{\rm drag}$ (the redshift and temperature of the Compton drag epoch).  The middle column is for the $\lc$ model with $T_0$ fixed at $2.7255\,$K, while the last column represents the parameter error on the $\lc{+}T_0$ model with the adoption of the FIRAS prior. The errors in the middle column (``$T_0$ fixed'') excluding those for $T_\ast$ and $T_{\rm drag}$ are presented in the final results of the Planck Collaboration~\cite{2018Planckparameter}.
}
\label{tab:CurrentFIRASError}
\end{table}

The parameter $\theta_{\rm MC}$ is a common alternative to $\theta_{\ast}$ used in the literature~\cite{2015Planckparameter, 2018Planckparameter, 2020ACTDR4}, which is much faster to compute. It is based on a numerical approximation for $\theta_{\ast}$, derived in Ref.~\cite{HuSmall-Scale1996}, with the assumption of $T_0$ being a constant. However, the error on $\theta_{\rm MC}$ increases substantially (and a large bias develops) when we vary $T_0$ as a free parameter without the FIRAS prior. The approximating formula for $\theta_{\rm MC}$ depends explicitly on $\bht$ and $\cht$ instead of $\bb$ and $\bc$, and therefore performs poorly when we start to vary $T_0$. Even when we only vary $T_0$ within the FIRAS prior, the $68\,\%$ error of $\theta_{\rm MC}$ still increases by about $20\,\%$ compared to the error with the fixed $T_0$ model (as seen in Table~\ref{tab:CurrentFIRASError}). However, the constraint on $\theta_{\ast}$, the accurate version of $\theta_{\rm MC}$ (as defined in Sec.~\ref{othercosmoparameters}), experiences no statistically significant change with or without fixing $T_0$. There is no compelling reason to continue using this approximate parameter $\theta_{\rm MC}$ in any case, given the currently available computing power.

A similar increase in the uncertainties can also be seen for $z_{\ast}$ and $z_{\rm drag}$ (the redshifts at the last-scattering epoch and the Compton-drag epoch, respectively) in Table~\ref{tab:CurrentFIRASError}. Since CMB power spectra directly constrain $T_{\ast}$ to an accuracy comparable to the FIRAS measurement (as established in Sec.~\ref{role_of_T0}) instead of $z_{\ast}$, the inclusion of the FIRAS error on $T_0$ will noticeably increase the error on $z_{\ast}$, considering the relation $z_{\ast}=T_{\ast}/T_0-1$. A similar argument to our discussion on the epoch of last scattering will apply to $T_{\rm drag}$ and $z_{\rm drag}$, which describe the time of decoupling of baryons from the photon background. Such an underestimation of parameter uncertainties by neglecting the FIRAS error will not occur if we choose to use $T_{\ast}$ and $T_{\rm drag}$ instead of $z_{\ast}$ and $z_{\rm drag}$ in the first place. It is therefore better to use temperature to describe the timeline of major events in the early Universe rather than redshift.

\subsection{Future CMB anisotropy experiments}\label{FutureImpactFIRASerror}

Even though the FIRAS error does not make a difference to the main parameter constraints or our actual cosmological model in the current generation of CMB experiments (and will not even for the next generation of ``Stage~4'' experiments), there is still the question of whether the FIRAS measurement will be sufficient for {\it all\/} the cosmological results of all future CMB experiments. In the rest of this section, we will assess the impact of the FIRAS prior on a CVL CMB experiment going out to high multipoles.  A similar question was addressed in Refs.~\cite{Hamann2008} and~\cite{barT}; however, Ref.~\cite{Hamann2008} incorrectly propagated the dipole error into the parameter analysis and only considered spectra up to $\ell_{\rm max} = 2500$.  Reference~\cite{barT} focused on a {\em Planck}-like experiment and also considered the effect of the cosmic variance of $T_0$; we will discuss this work further in Appendix~\ref{sec:inhomogeneous}. 

\begin{table}[htbp!]
\setlength{\tabcolsep}{10pt}
\begin{tabular}{l c c}
\hline
\hline
\noalign{\vskip 1pt}
Parameter& $T_0$ fixed& FIRAS prior\\
\hline
\noalign{\vskip 2pt}
$\Omega_{\rm b} h^2$& $1.12\times 10^{-5}$& $1.85\times 10^{-5}$\\
$\Omega_{\rm c} h^2$& $2.09\times 10^{-4}$& $2.22\times 10^{-4}$\\
\hline
\end{tabular}
\caption{Estimated uncertainties for $\Omega_{\rm b} h^2$ and $\Omega_{\rm c} h^2$ from a 7-parameter Fisher-matrix calculation using the lensed $TT$, $TE$, and $EE$ power spectra, assuming that we are in the cosmic-variance limit. The second column is for $T_0$  fixed at 2.7255\,K, while the third column is for adoption of the FIRAS prior. In this calculation, we go to $\ell_{\rm max}=3000$ for the $TT$ power spectrum and $\ell_{\rm max}=6000$ for the $TE$ and $EE$ power spectra. More details are given in Appendix~\ref{sec:method}.}
\label{tab:3000-6000result}
\end{table}

\begin{figure}[htbp!]
\centerline{\includegraphics[width=\hsize]{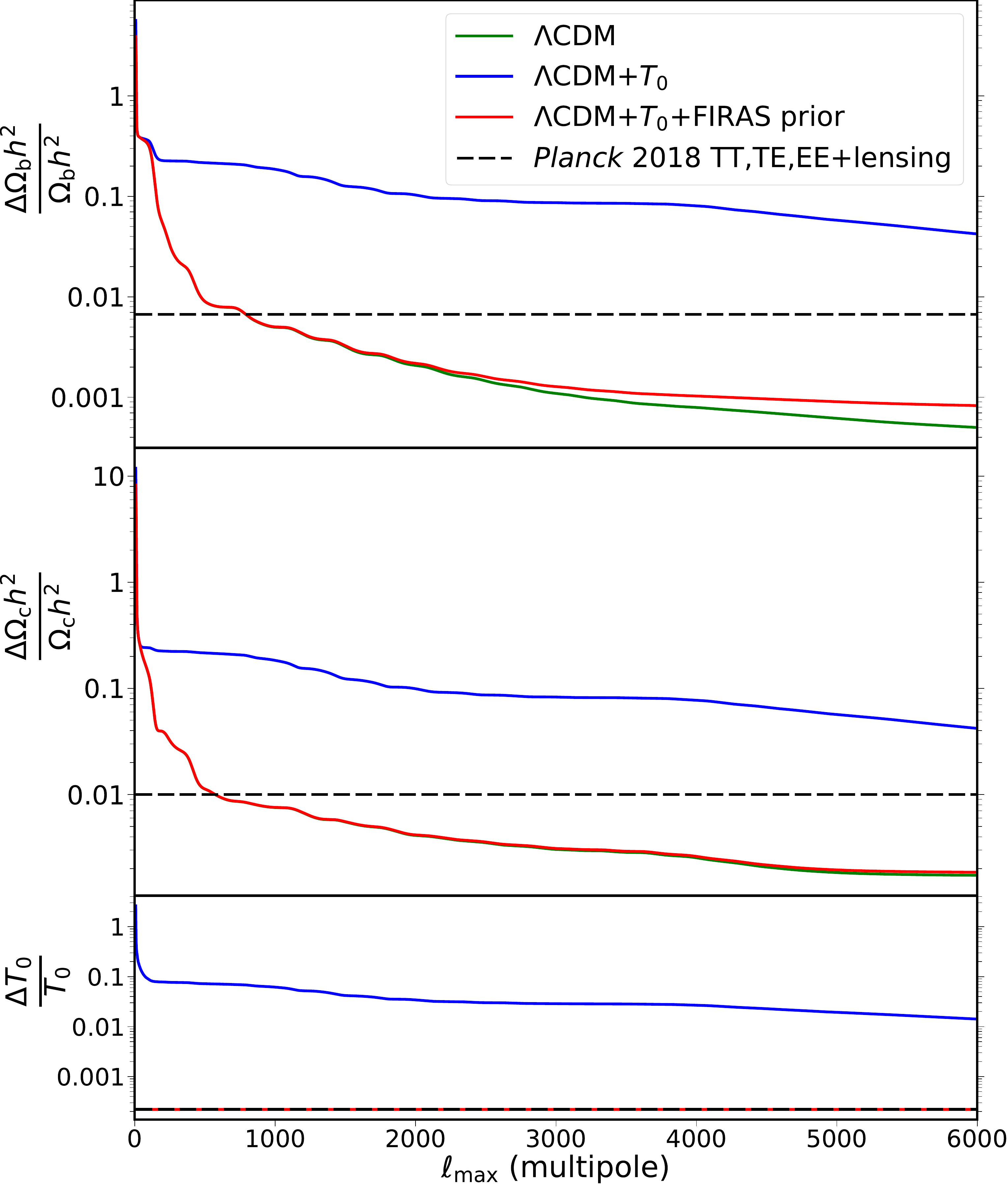}}
\caption{Estimated Fisher-matrix uncertainties for $\Omega_{\rm b} h^2$, $\Omega_{\rm c} h^2$, and $T_0$ as a function of the maximum $\ell$, using lensed $TT$, $TE$, and $EE$ power spectra under the assumption of the cosmic-variance limit, as described in Appendix~\ref{sec:method}. The green and blue curves show the constraints for the free 6- and 7-parameter models, respectively. The red curve shows the constraints for a 7-parameter $\lc$+$T_0$ model while imposing the FIRAS prior (i.e., following Eq.~\ref{eq: FIRAS_Error}) on $T_0$. The horizontal black dashed lines give the uncertainties of $\bht$ and $\cht$ derived from the actual 2018 {\em Planck} TT,TE,EE+lensing likelihood, along with the uncertainty from the FIRAS $T_0$ measurement.}
\label{fig:temp_firas}
\end{figure}

We now examine the effects of including the FIRAS prior on the constraints of the standard-$\lc$ parameters $\{\Omega_{\rm b} h^2,\Omega_{\rm c} h^2, \theta_{\ast}, A_{\rm s}, n_{\rm s}, \tau\}$ using ideal CVL CMB data. The uncertainties for $\{\theta_{\ast}, A_{\rm s}, n_{\rm s},\tau\}$ are different by less than $0.1\,\%$ in the two settings; hence these can be considered to be unchanged for all practical purposes. However, as shown in Table~\ref{tab:3000-6000result} and Fig.~\ref{fig:temp_firas}, the constraints for $\bht$ and $\cht$ are noticeably different, with $\bht$ showing the greatest change among all parameters. Using a fixed temperature (i.e., ignoring the FIRAS uncertainty) would lead us to underestimate the error for $\bht$ by around $50\,\%$ and the error for $\cht$ by around $10\,\%$ for $\ell_{\rm max}\simeq6000$. Therefore, for an ideal future experiment, the FIRAS uncertainty {\it does\/} have an effect on the standard-model 6-parameter constraints; nevertheless, such differences are modest in size, and in practice any differences would be even smaller because of the inclusion of realistic noise from the instrument and foreground residuals.  We can conclude that imposing the FIRAS prior on $T_0$ will be sufficient for analysing all CMB experiments in the near future, until such time as we can approach the cosmic-variance limit out to multipoles of many thousands.  It is therefore generally safe to continue the practice of treating $T_0$ as a constant, without any measurable impact on the derived parameters in standard cosmological models. We also checked the effect of adding the lensing reconstruction spectrum $C_{\ell}^{\rm \phi\phi}$ in the Fisher matrix calculation, and our results still hold (although the uncertainties are slightly smaller).

When we use the alternative parameter set with $\bb$ and $\bc$, the parameter constraints from the 7-parameter $\lc{+}T_0$ model with a FIRAS prior are essentially the same as those from the 6-parameter $\lc$ model under CVL conditions forecasted using a Fisher matrix formalism, as shown through the perfectly overlapping green and red curves in Fig.~\ref{fig:bbbcequivalence}. Since $\bht\propto\bb T_0^3$ and $\cht\propto\bc T_0^3$, then the relative errors on $\bht$ and $\cht$ with the FIRAS prior are simply given by those with fixed $T_0$ added in quadrature with 3 times the FIRAS uncertainty.
The FIRAS uncertainty plays a significant role when the constraints on the parameters $\bht$ (or $\bb$) and $\cht$ (or $\bc$) approach the FIRAS relative error. This can be seen in Fig.~\ref{fig:temp_firas}, where the constraint on $\bht$ begins to deviate at around $\ell_{\rm max}\simeq2000$, with the relative error approaching $10^{-3}$. Because $\bht$ is constrained by high-$\ell$ data, the FIRAS uncertainty leads to a more significant effect and eventually causes the $50\,\%$ underestimation of the error at $\ell_{\rm max}\simeq6000$. The errors on $\cht$ for the two models also begin to slightly differ in Fig.~\ref{fig:temp_firas} as the relative error approaches $10^{-3}$, for $\ell_{\rm max}$ extending to 6000. However, since $\cht$ is not as well constrained by CMB anisotropy data as $\bht$, the FIRAS uncertainty impacts the constraints on $\bht$ much more than those on $\cht$. To summarize, the FIRAS uncertainty will have a small (but non-negligible) effect only when $\bht$ and $\cht$ are constrained to about the $0.1\,\%$ level in $\lc$. 

\begin{table}[htbp!]
\setlength{\tabcolsep}{10pt}
\begin{tabular}{l c c}
\hline
\hline
\noalign{\vskip 1pt}
Parameter& $T_0$ fixed& FIRAS prior\\
\hline
\noalign{\vskip 2pt}
100$\theta_{\rm MC}$& $6.0\times10^{-5}$ & $2.4\times10^{-4}$\\
100$\theta_{\ast}$ & $6.0\times10^{-5}$ & $6.0\times10^{-5}$\\
$z_{\ast}$ & $0.021$ & $0.20$ \\
$T_{\ast}$ [K] & $0.056$ & $0.056$\\
$z_{\rm drag}$ &  $0.033$ & $0.24$\\
$T_{\rm drag}$ [K] & $0.090$ & $0.090$\\
\hline
\end{tabular}
\caption{Estimated uncertainties for some derived parameters from a 7-parameter Fisher-matrix calculation using the lensed $TT$, $TE$, and $EE$ power spectra, assuming that we are in the cosmic-variance limit. These estimates are the absolute uncertainties on the parameter (instead of the percentage error). This table is the CVL forecast version of Table~\ref{tab:CurrentFIRASError}. In this calculation, we go to $\ell_{\rm max}=3000$ for the $TT$ power spectrum and $\ell_{\rm max}=6000$ for the $TE$ and $EE$ power spectra. More details are given in Appendix~\ref{sec:method}.}
\label{tab:FutureFIRASError-derived}
\end{table}
Following Sec.~\ref{CurrentImpactFIRASerror}, we also forecast the uncertainties on the derived cosmological parameters listed in Table~\ref{tab:CurrentFIRASError}, assuming ideal CMB data in the 6-parameter $\Lambda {\rm CDM}$ model. As shown in Table~\ref{tab:FutureFIRASError-derived}, in the cosmic-variance limited setting, the inclusion of the FIRAS error for $T_0$ increases the uncertainties on $\theta_{\rm MC}$, $z_{\ast}$, and $z_{\rm drag}$ substantially compared to Table~\ref{tab:CurrentFIRASError} with \textit{Planck} data. In fact, ignoring the FIRAS error underestimates the uncertainties by almost an order of magnitude. This provides compelling evidence that for future CMB analysis we should adopt $\theta_{\ast}$, $T_{\ast}$, and $T_{\rm drag}$ as more appropriate parameters that are not sensitive to the change of $T_0$, compared with their currently-used counterparts.

\begin{table}[htbp!]
\setlength{\tabcolsep}{10pt}
\begin{tabular}{l c c}
\hline
\hline
\noalign{\vskip 1pt}
Parameter& $T_0$ fixed& FIRAS prior\\
\hline
\noalign{\vskip 2pt}
$\Omega_{\rm b} h^2$& $1.13\times 10^{-5}$& $1.85\times 10^{-5}$\\
$\Omega_{\rm c} h^2$& $5.29\times 10^{-4}$& $5.35\times 10^{-4}$\\
$\Omega_{K}$& $1.29\times 10^{-3}$& $1.29\times 10^{-3}$\\
\hline
\end{tabular}
\caption{Estimated uncertainties for $\Omega_{\rm b} h^2$, $\Omega_{\rm c} h^2$, and $\Omega_{K}$ from an 8-parameter $\lc{+}\Omega_{K}{+}T_0$ model Fisher-matrix calculation using the lensed $TT$, $TE$, and $EE$ power spectra, assuming that we are in the cosmic-variance limit. The calculation setting is the same as for Table~\ref{tab:3000-6000result}, with more details given in Appendix~\ref{sec:method}.} 
\label{tab:3000-6000result-omk}
\end{table}

We also examine the effects of including the FIRAS prior on the parameter constraints for some models that are extensions to 6-parameter $\Lambda$CDM. In general, the FIRAS prior has negligible impact on these extended parameters in an ideal CVL anisotropy experiment.  To illustrate this we take $\Omega_K$ as an example of such an extended parameter. As seen in Table~\ref{tab:3000-6000result-omk}, the inclusion of the FIRAS uncertainty has a similar impact on $\bht$ and $\cht$ for the $\lc{+}\Omega_K$ model compared to the results in Table~\ref{tab:3000-6000result} for $\lc$; this means that our previous discussion on the impact of the FIRAS uncertainty on the $\lc$ model still holds in this case. Varying $\Omega_K$ as a free parameter increases the error in a substantial way only for $\cht$, due to the degeneracy between the matter density and curvature. However, including the FIRAS uncertainty does not impact the error on $\Omega_K$, since the constraint for $\Omega_K$ obtained in a CVL CMB anisotropy experiment is still far from the relative accuracy of the FIRAS measurement. Including a \textit{Euclid}-like measurement of the BAO scale (see Appendix~\ref{sec:method} for details) further strengthens the parameter constraints, but the FIRAS error still has no impact on the accuracy of $\Omega_K$. Because the FIRAS measurement is so precise, the uncertainty on $T_0$ will generally not be a concern for future CMB anisotropy experiments placing constraints on the $\lc{+}\Omega_K$ model, or on other common extensions to 6-parameter $\Lambda$CDM.

\subsection{Future LSS measurements}\label{FutureLSSFIRAS}
For ideal CMB anisotropy experiments, we saw that the FIRAS uncertainty starts to inhibit parameter constraints when the errors of $\bht$ and $\cht$ approach the $0.1\,\%$ level. For upcoming LSS surveys, the uncertainties on some cosmological parameters might also approach this precision. For example, with the combined cosmological probes from galaxy clustering and weak lensing in an optimistic scenario, \textit{Euclid} is projected to constrain several cosmological parameters (including $\sigma_{8}$, $h$, $\Omega_{\rm b}$, and $\Omega_{\rm m}$) to well below the percent level~\cite{19EuclidForecast}, where the present-day uncertainty on $T_0$ could actually play a role. A more rigorous assessment is needed to determine whether the uncertainty of $T_0$ will actually matter in practice for such LSS surveys. 
 
\section{Conclusions}
\label{conclusions}

In this paper, we have investigated the effect of the parameter $T_0$ on CMB anisotropies.  In order to clarify the role of $T_0$ in calibration, we advocate the use of dimensionless (i.e., $\Delta T/T$) units to measure, analyse, and present the CMB dipole, as well as the higher multipoles of the CMB sky. When calibrating with the orbital dipole, the dimensionless CMB fluctuations can be measured independently of our knowledge of $T_0$, even in the presence of foregrounds. The value $T_{\rm c}=\overline{T}_{0, \rm F}=2.7255\,{\rm K}$ is merely a particular choice of calibration constant for unit-conversion purposes. Even though dimensionless units and $T_{\rm c}$-calibrated temperature units are physically equivalent, it is in principle better to use dimensionless quantities for CMB data, since this avoids unnecessary confusion between $T_0$ and $T_{\rm c}$ and prevents people from folding additional temperature uncertainty into the cosmological results. Since the CMB power spectra are naturally dimensionless in theory and are also measured directly in $\Delta T/T$ units, we should go back to the tradition of using dimensionless units, as theorists did in the earliest discussions of CMB anisotropies. 

We have also given an overview of the role of $T_0$ as a cosmological parameter, building on the work of Refs.~\cite{T0H0} and~\cite{openhotteruniverse}. As a cosmic clock, $T_{\gamma}$ indicates the change of absolute energy scale as the Universe evolves, while $T_0=T_{\gamma}(z=0)$ characterizes the amount of expansion after recombination, a process that is relatively well determined by the primary anisotropies. Clarifying some previous confusion on the role of $T_0$, we emphasize that the monopole temperature $T_0$ does not influence the physics of the early Universe, in particular recombination, when holding fixed the parameters $\bb$ and $\bc$, which are physically relevant at that time and are well constrained. As a result, the constraining power of $T_0$ from CMB power spectra only comes from late-time effects, in particular the ISW effect and gravitational lensing, with both the lensing smoothing effect on the 2-point functions and the lensing reconstruction spectrum itself providing comparable constraining power. Current CMB anisotropy data give only a weak constraint on $T_0$, which leads to various degrees of degeneracy between $T_0$ and other cosmological parameters, including $H_0$.  However, employing such degeneracies to address cosmic tensions ignores the fact that it would require excursions hundreds of times the size of the FIRAS uncertainty to make a substantial difference.  This means we would have to disregard current reliable knowledge of $T_0$, derived not just from {\it COBE}-FIRAS, but from many other experiments. Using a Fisher-matrix analysis for an ideal future experiment, we estimate that CMB data alone can at best constrain $T_0$ to the $1\,\%$ level, which is still 2 orders of magnitude less constraining than the current local temperature measurements. Adding \textit{Euclid}-like BAO results to the parameter forecasts improves the $T_0$ constraint by another factor of 4. It is in principle possible to obtain an independsent measurement of $T_0$ with a precision that could approach that of FIRAS by combining future CMB and LSS data.

Despite the pedagogical discussion of treating $T_0$ as a free variable, we are in the situation where it is sufficient to use the FIRAS measurement as a prior in order to provide any realistic assessment of the impact of $T_0$ on parameters extracted from CMB power spectra. The FIRAS uncertainty is so small that we can say it will generally have negligible impact on the main cosmological results coming from current and next-generation CMB anisotropy experiments. Hence adopting the central FIRAS value as $T_0$ will be sufficient for any currently proposed experiment. Nevertheless, it is important to keep in mind that neglecting the error of the FIRAS measurement will noticeably underestimate the uncertainties for several occasionally-quoted derived parameters, even in current CMB experiments, although such effects can be mitigated by choosing alternative derived parameters that are less sensitive to the change of $T_0$. As experimental capabilities improve, the FIRAS uncertainty will eventually impact the constraints on $\bht$ and $\cht$ for the main cosmological parameters as well. In the CVL situation, we will underestimate the uncertainty on $\Omega_{\rm b} h^2$ by $50\,\%$ and on $\Omega_{\rm c} h^2$ by $10\,\%$ (out to $\ell_{\rm max}=6000$ for polarization and 3000 for temperature) if we ignore the FIRAS uncertainty on $T_0$. This shows that if we ultimately want to extract all available information from the CMB power spectra measured to multipoles $\ell\simeq5000$, then we will indeed need a better determination of $T_0$ than is currently available. Our work thereby provides another motivation for the proposed future CMB spectral distortion experiments \cite{17PIXIEspectraldistortion,19SpectralDistortion1,19SpectralDistortion2}, which will further improve the accuracy of our measurement for $T_0$ compared to FIRAS.

\section*{Acknowledgements}
We thank Jens Chluba and Hans Kristian Eriksen for useful discussions. This research was supported by the Natural Sciences and Engineering Research Council of Canada. Computing resources were provided by Compute Canada/Calcul Canada (\url{www.computecanada.ca}). 
Parts of this paper are based on observations obtained with \textit{Planck} (\url{www.esa.int/Planck}), an ESA science mission with instruments and contributions directly funded by ESA Member States, NASA, and Canada.
This paper made use of the codes
\texttt{CAMB} (\url{camb.readthedocs.io/en/latest/}),
\texttt{Cobaya} (\url{cobaya.readthedocs.io/en/latest/}), and \texttt{Recfast} (\url{www.astro.ubc.ca/people/scott/recfast.html}).

\appendix
\section{Fisher-matrix method}\label{sec:method}

The Fisher-matrix formalism is a well established tool for forecasting parameter constraints in cosmology~\cite{Fisher1922,TTH97,09StatinCosmology,cosmologyfishermatrix}.  We use this approach to assess the impact of varying $T_0$ on CMB power spectra in both realistic and cosmic-variance-limited (CVL) conditions. For any observed data vector $\mathbf{x}$ and model parameters $\mathbf{p}$, with likelihood function $L(\mathbf{x};\mathbf{p})$, the Fisher information matrix is defined as 
\begin{equation}
{\sf F}_{ij}=-\left\langle\frac{\partial^2\ln L}{\partial p_i \partial p_j}\right\rangle. 
\end{equation}
By the Cramer-Rao inequality, the variance of an unbiased estimator for a parameter $p_i$ from the data has a lower bound $({\sf F}^{-1})_{ii}$, to which the maximum likelihood estimator approaches asymptotically. Therefore, the Fisher matrix gives an estimate for the approximate error bars achievable from an experiment.

\subsection{Fisher matrix for CMB}
The Fisher matrix for CMB temperature and polarization anisotropies is~\cite{CMBfishermatrix,SantosLensingFisherMatrix}
\beq
{\sf F}_{ij}^{\rm CMB}=\sum_{\ell=2}^{\ell_{\rm max}}\sum_{X,Y}\frac{\partial C_{\ell}^{X}}{\partial p_i}({\rm Cov}_\ell)_{XY}^{-1}\frac{\partial C_{\ell}^Y}{\partial p_j},
\label{eq:fisher matrix}
\eeq
assuming Gaussian primordial perturbations and Gaussian noise. $C_{\ell}^{X}$ represents the power of the $\ell$th multipole for $X={\rm T,E,C}$, which stand for either the lensed or unlensed $TT$, $EE$, or $TE$ power spectra, respectively. We neglect the $BB$ power spectrum, since the detection of the primordial signal is not guaranteed and, as we have confirmed, adding the $BB$ power spectrum does not significantly change our results.\footnote{In fact the $BB$ spectrum does not add significant constraining power to either the 6-parameter $\lc$ or 7-parameter $\lc{+}T_0$ models, except by noticeably improving the constraints on the dark matter density $\Omega_{\rm c} h^2$ by around a factor of 2 at $\ell_{\rm max}\simeq2000$ for the 6-parameter $\lc$ model. However, the differences between the constraints for $\Omega_{\rm c} h^2$ with or without the $BB$ spectrum become negligible by $\ell_{\rm max}\simeq5000$. Most of the lensing information contained in the $BB$ power spectrum is also contained in the $EE$ power spectrum in the CVL case.} We use \texttt{CAMB} for all the CMB power spectrum calculations, with \texttt{Recfast} chosen as the recombination code; we checked that selecting other recombination codes does not impact any of our results.  The symmetric covariance matrix ${\rm Cov}_\ell$ has elements
\begin{align}
    \label{cov1}
    ({\rm Cov}_\ell)_{\rm TT}& = \frac{1}{\mathcal{N}}\left(C_{\ell}^{\rm T}+\omega_{\rm T}^{-1}B_{\ell}^{-2}\right)^2,\\
    ({\rm Cov}_\ell)_{\rm EE}& = \frac{1}{\mathcal{N}}\left(C_{\ell}^{\rm E}+\omega_{\rm P}^{-1}B_{\ell}^{-2}\right)^2,\\
    ({\rm Cov}_\ell)_{\rm CC}& = \frac{1}{2\mathcal{N}}\left[\left(C_{\ell}^{\rm C}\right)^2\right.\nonumber\\
    & +\left(C_{\ell}^{\rm T}+\omega_{\rm T}^{-1}B_{\ell}^{-2}\right)\left(C_{
    \ell}^{\rm E}+\omega_{\rm P}^{-1}B_{\ell}^{-2}\right)\Big],\\
    ({\rm Cov}_\ell)_{\rm TE}& = \frac{1}{\mathcal{N}}\left(C_{\ell}^{\rm C}\right)^2,\\
    ({\rm Cov}_\ell)_{\rm TC}& =  \frac{1}{\mathcal{N}}C_{\ell}^{\rm C}\left(C_{\ell}^{\rm T}+\omega_{\rm T}^{-1}B_{\ell}^{-2}\right),\\
    ({\rm Cov}_\ell)_{\rm EC}& = \frac{1}{\mathcal{N}}C_{\ell}^{\rm C}\left(C_{\ell}^{\rm E}+\omega_{\rm P}^{-1}B_{\ell}^{-2}\right),\label{covlast}
\end{align}
where we define 
$\mathcal{N}=f_{\rm \rm sky}(2\ell+1)/2$, and we take the \rm \rm sky coverage fraction $f_{\rm \rm sky}=0.7$ throughout the paper (so in this sense our estimates have some degree of realism, even though we neglect any complications related to foreground removal). The beam window function $B_{\ell}^2$ is assumed to be Gaussian with $B_{\ell}^2=\exp[-\ell(\ell+1)\theta_{\text{beam}}^2/(8\ln2)]$, where $\theta_{\rm beam}$ is the full-width, half maximum (FWHW) of the beam.  The quantities $\omega_{\rm T}$ and $\omega_{\rm P}$ are inverse squares of the detector noise level per steradian for temperature and polarization, respectively; these can be determined by $\omega_{\rm T, P}=(\theta_{\text{beam}}\sigma_{\rm T,P})^{-2}$, where $\sigma_{\rm T}$ and $\sigma_{\rm P}$ are the noise in units of $\mu {\rm K}$ per FWHM beam size. For convenience, we use a single channel as a simple approximation for modelling the \textit{Planck} results, taking values from Ref.~\cite{CMBfishermatrix}, namely $\theta_{\text{beam}}=5.5$, $\sigma_{\rm T}=11.7\,\mu{\rm K}$, and $\sigma_{\rm P}=24.3\,\mu{\rm K}$.  We have verified that this noise specification provides an excellent forecast for the current \textit{Planck} results in~\cite{2018Planckparameter}.  We sum $\ell$ from 2 to 2500 with the above beam size and white noise specification, which we used to generate Fig.~\ref{fig:planck_error_T0}. Since the noise here is specified in $\mu$K, we use \eq({\ref{eq:C_l unit}}) to convert the dimensionless $C_{\ell}$ into $\mu {\rm K}^2$ units, as described in Sec.~\ref{sec:calibration} for CMB power spectra.  Ideally, the noise should be specified as dimensionless as well and will have the same calibration as the signal; however, in practice even if the noise was calibrated differently than the signal this would make negligible difference (since changing the calibration choice will at most lead to a small difference in the noise level, and hence this is effectively a higher-order correction for parameter forecasts).

In the CVL case, we simply remove the noise term $\omega_{\rm T, P}^{-1}B_{\ell}^{-2}$ from the covariance matrix (Eqs.~\ref{cov1}--\ref{covlast}) to obtain the correct Fisher matrix, working entirely with dimensionless $C_{\ell}$. \textit{Planck} has already measured most of the information from the $TT$ power spectrum out to $\ell\simeq2000$, and with
foregrounds it seems unrealistic to push beyond $\ell\simeq3000$ for $TT$~\cite{ScottInformationCMB}. However, since the foregrounds
from galaxies and galaxy clusters are very weakly polarized, we should be able to measure the primary polarization anisotropies out to $\ell\simeq5000$ and perhaps even higher~\cite{ScottInformationCMB,ProspectDelensingInflation}. Therefore, in our noise-free experiment, we take $\ell_{\rm max}=3000$ for $TT$ and $\ell_{\rm max}=6000$ for the $TE$ and $EE$ power spectra in Eq.~(\ref{eq:fisher matrix}). The $C_{\ell}$ power spectra become non-Gaussian at $\ell\simeq5000$ due to the effects of lensing on small scales, so the Fisher forecast will not be exactly accurate there. However, the Fisher forecast still gives an approximate estimate of the error. Instead of providing an exact error forecast, we are mostly interested in the general effects of the FIRAS error on cosmological parameter constraints, so the use of Fisher matrix is sufficient for our purpose. Figures~\ref{fig:bbbcequivalence} and \ref{fig:temp_firas} are obtained using these CVL settings.

To combine independent likelihood functions, we can add the corresponding Fisher matrix of each likelihood function to find the total Fisher matrix, and the FIRAS prior is indeed independent of the CMB likelihood function. For a normal distribution for $T_{0,{\rm F}}$ with standard deviation $\sigma_{\rm F}$ 
that characterizes the FIRAS measurement in Eq.~(\ref{eq: FIRAS_Error}), where $\sigma_{\rm F}=0.0006\,{\rm K}$ is known, the $T_0$--$T_0$ term of the Fisher matrix is $1/\sigma_{\text{F}}^2$; we therefore just add $1/\sigma_{\text{F}}^2$ to the $T_0$--$T_0$ term in the CMB Fisher matrix calculated from Eq.~(\ref{eq:fisher matrix}) to account for the FIRAS prior in the $\lc{+}T_0$ model.

To study the constraining power of the lensing reconstruction spectrum $C_{\ell}^{\rm \phi\phi}$ on $T_0$ in the CVL setting, we include $C_{\ell}^{\rm \phi\phi}$ in \eq({\ref{eq:fisher matrix}}). The additional terms in the symmetric covariance matrix are given by~\cite{SantosLensingFisherMatrix}
\begin{align}
    \label{lenscov1}
    ({\rm Cov}_\ell)_{{\rm T}\phi}& = \frac{1}{\mathcal{N}}\left(C_{\ell}^{{\rm T}\phi}\right)^2,\\
    ({\rm Cov}_\ell)_{{\rm E}\phi}& = \frac{1}{\mathcal{N}}\left(C_{\ell}^{{\rm E}\phi}\right)^2,\\
    ({\rm Cov}_\ell)_{{\rm C}\phi}& = \frac{1}{\mathcal{N}}C_{\ell}^{{\rm T}\phi}C_{\ell}^{{\rm E}\phi},\\
    ({\rm Cov}_\ell)_{\phi\phi}& = \frac{1}{\mathcal{N}}\left(C_{\ell}^{\phi\phi}\right)^2.
    \label{lenscovlast}
\end{align}
We take the lensing reconstruction spectrum to extend to $\ell_{\rm max}=1000$ in \eq({\ref{eq:fisher matrix}}). Here the CVL condition refers to the ideal no-noise assumption for the measurement the of CMB lensing reconstruction spectrum $C_{\ell}^{\phi\phi}$. We exclude the actual reconstruction noise for simplicity. Figure~\ref{fig:lensing_T0} is obtained with the inclusion of $C_{\ell}^{\rm \phi\phi}$ in the Fisher matrix, assuming CVL settings.

\subsection{Fisher matrix for BAO}

\begin{table*}[htbp!]
\begin{tabular}{p{0.13\textwidth}<{\centering}p{0.05\textwidth}<{\centering}p{0.05\textwidth}<{\centering}p{0.05\textwidth}<{\centering}p{0.05\textwidth}<{\centering}p{0.05\textwidth}<{\centering}p{0.05\textwidth}<{\centering}p{0.05\textwidth}<{\centering}p{0.05\textwidth}<{\centering}p{0.05\textwidth}<{\centering}p{0.05\textwidth}<{\centering}p{0.05\textwidth}<{\centering}p{0.05\textwidth}<{\centering}p{0.05\textwidth}<{\centering}p{0.05\textwidth}<{\centering}p{0.05\textwidth}<{\centering}} 
\hline
\hline
\noalign{\vskip 1pt}
$z$ & 0.65 & 0.75 & 0.85 & 0.95 & 1.05 & 1.15 & 1.25 & 1.35 & 1.45 & 1.55 & 1.65 & 1.75 & 1.85 & 1.95 & 2.05\\ 
\hline
\noalign{\vskip 2pt}
$(\Delta y/y)^\perp$ [$\%$] & 1.23 & 0.83 & 0.74 & 0.71 & 0.70 & 0.70 & 0.70 & 0.73 & 0.78 & 0.87 & 1.01 & 1.23 & 1.61 & 2.32 & 5.32\\
$(\Delta y/y)^\parallel$ [$\%$] & 1.89 & 1.42 & 1.27 & 1.19 & 1.14 & 1.12 & 1.10 & 1.11 & 1.16 & 1.24 & 1.40 & 1.64 & 2.07 & 2.90 & 6.39\\
\noalign{\vskip 1pt}
\hline
\end{tabular}
\caption{Estimated errors for BAO scale measurements in a $\textit{Euclid}$-like large-scale structure survey (see Ref.~\cite{14BAONeutrinoForecast} for further details). The first row shows the central redshift of each observational bin. The other rows give the estimated percentage error $\Delta y/y$ on the transverse and radial BAO scales, where $y^\perp(z) = D_{\rm A}(z)/r_{\rm s}$ and $y^\parallel(z) = H(z)r_{\rm s}$, respectively.
}
\label{tab:EuclidBAOforecast}
\end{table*}

To break the geometric degeneracy and further constrain the late-time expansion, it is common practice to combine CMB data with BAO measurements for estimating cosmological parameters. BAO surveys generally use galaxy clustering to measure the transverse and radial scales of the sound horizon at recombination. The uncertainty in BAO experiments can be given as the errors on the transverse and radial BAO scale parameters $y^\perp(z) =D_{\rm A}(z)/r_{\rm s}$ and $y^\parallel(z)=H(z)r_s$, respectively.  We use $\Delta y^\perp$ and $\Delta y^\parallel$ to represent the errors associated with their measurement. 
For simplicity, we assume that the errors in $y^\perp$ and $y^\parallel$ are uncorrelated with each other and that there are no correlations between different redshift bins.  In this case, the covariance matrix is diagonal, and the full Fisher matrix for BAO simplifies to~\cite{14QuintessenceModelEuclid}
\begin{align}
{\sf F}_{ij}^{\rm BAO}&=\sum_{i}\frac{1}{{(\Delta y^\perp_i})^2}\frac{\partial y^\perp(z_i)}{\partial p_i}\frac{\partial y^\perp(z_i)}{\partial p_j}\nonumber\\
&+\sum_{i}\frac{1}{{(\Delta y^\parallel_i})^2}\frac{\partial y^\parallel(z_i)}{\partial p_i}\frac{\partial y^\parallel(z_i)}{\partial p_j},
\label{eq:BAOFisherMAtrix}
\end{align}
where the sums run over the observational bins at different redshifts. We take the central redshift values of each bin to calculate $y^\perp(z)$ and $y^\parallel(z)$ using \texttt{CAMB}. The errors forecast for a \textit{Euclid}-like experiment are given explicitly in Table~\ref{tab:EuclidBAOforecast}, and are derived from Ref.~\cite{14BAONeutrinoForecast}, using a method based on the work of Refs.~\cite{03SeoBAO} and~\cite{07SeoBAO}.  To combine the forecasts coming from CMB and BAO measurements, we add the Fisher matrix results from each experiment. These combined CMB and BAO forecasts are used to estimate future parameter constraints on $T_0$ in Sec.~\ref{constraining T0} and $\Omega_K$ in Sec.~\ref{FutureImpactFIRASerror}.  

\section{Temperature definitions}
\label{sec:tempdefs}
There are several temperature-related quantities referred to in this paper.
In order to distinguish between them,
Table~\ref{tab:tempdefs} provides definitions and values (where appropriate) for various temperatures.

\begin{table*}[htbp!]
\setlength{\tabcolsep}{4pt}
\begin{tabular}{l p{11cm} c}
\hline
\hline
\noalign{\vskip 1pt}
Parameter & Definition & Value\\
\hline
\noalign{\vskip 1pt}
$\Delta T/T$& Dimensionless CMB anisotropy. We use this schematically, with no
precise meaning for the denominator.& \dots \\
\noalign{\vskip 1pt}
$\Delta T_{\rm RJ}$& Rayleigh-Jeans temperature fluctuation, $\Delta I_{\nu}c^2/2k\nu^2$.& \dots \\
\noalign{\vskip 1pt}
$T_0$& Present-day temperature of the photon background, or the monopole term in the spherical-harmonic expansion of the CMB sky.& \dots \\
\noalign{\vskip 1pt}
$T_{0, \rm F}$& Current CMB monopole temperature measurement, derived by {\it COBE}-FIRAS combined with other experiments.& $(2.7255\pm0.0006)\,$K\\
\noalign{\vskip 1pt}
$\overline{T}_{0, \rm F}$& Central value for current monopole temperature measurement (i.e., the FIRAS measurement), often used as the calibration temperature.& $2.7255\,$K \\
\noalign{\vskip 1pt}
$T_{\rm c}$& Arbitrary fixed (errorless) calibration temperature, used to convert dimensionless power spectra to temperature units. Often set to $\overline{T}_{0, \rm F}$.& \dots \\
\noalign{\vskip 1pt}
$T_\gamma$, $T_\gamma(z)$& Photon temperature at redshift $z$.  $T_0=T_\gamma(z\,{=}\,0)$.& \dots \\
\noalign{\vskip 1pt}
$T_\ast$& Temperature at the epoch of last scattering, fixed by the physics of recombination.  $T_\ast=T_\gamma(z_\ast)$.  Value from this work and Ref.~\cite{2018Planckparameter}.& ($2973.27\pm0.67)\,$K \\
\noalign{\vskip 1pt}
$T_{\rm e}$& Kinetic temperature of the electrons.& \dots \\
\noalign{\vskip 1pt}
$T_{\rm drag}$& Temperature at the Compton drag epoch. $T_{\rm drag}=T_\gamma(z_{\rm drag})$.  Value from this work and Ref.~\cite{2018Planckparameter}.& $(2891.60\pm 0.78)\,$K\\
\hline
\end{tabular}
\caption{Different temperatures and definitions used throughout this paper. $T_*$ and $T_{\rm drag}$ values are taken from Table~\ref{tab:CurrentFIRASError}.}
\label{tab:tempdefs}
\end{table*}

\section{Inhomogeneity and \texorpdfstring{\boldmath{$T_0$}}{T0}} \label{sec:inhomogeneous}

Throughout this paper we have implicitly treated $T_0$ as a background 
cosmological parameter, i.e., as spatially constant.  In reality the 
gravitational redshifts and blueshifts due to structure change this picture, 
resulting in spatial as well as time dependence for the CMB temperature.  
A straightforward example of this is that the gravitational potential at the surface of 
the Earth, compared to the potential at a great distance, increases the 
CMB temperature on the ground by a factor $1 + GM_\oplus/R_\oplus c^2 
\simeq 1 + 7\times10^{-10}$ over the distant value.  While this is a
negligible amount, structure on large scales is expected to perturb $T_0$ at the $10^{-5}$ 
level\footnote{Metric perturbations caused by clusters and superclusters of galaxies are at the $10^{-5}$ level, this being one of the ``six numbers'' described by Martin Rees~\cite{SixNumbers}, which makes peculiar motions with kinetic energies that have $v^2/c^2$ of order $10^{-5}$.}~\cite{barT}, which is only of order a tenth the FIRAS uncertainty.  Therefore, when we consider the ultimate effect of the $T_0$ uncertainty with experiments that supersede FIRAS, it will be important to take into account the inhomogeneity of the Universe.

The perturbations of the CMB temperature over constant-time slices of the 
spacetime result in a cosmic variance, $C_0$, in $T_0$ that was first studied for 
particular slices (i.e., in a particular gauge) in Ref.~\cite{gauging}.  
In Ref.~\cite{barT} the result of a calculation of $C_0$ in comoving 
gauge~\cite{by21} was reported and found to have 
a negligible effect on the uncertainties of the other cosmological parameters 
relative to that of the FIRAS uncertainty.  Note, however, that the choice of gauge in these 
studies was arbitrary, and can lead to a zero, order $10^{-5}$, or divergent result for 
$C_0$~\cite{gauging}.  For an eventual successor to FIRAS,
we note that the effect of super-Hubble fluctuations in $T_0$ should be 
irrelevant insofar as the effect on the cosmological parameters within our 
Hubble volume is concerned.  We also point out that the perturbations due 
to sub-Hubble structure could, in principle, be taken into account and 
corrected for by mapping that structure.  So ultimately we do not expect the 
cosmic variance of $T_0$ to have an important effect on the determinations of 
the other parameters.

In further work related to the effect of inhomogeneity on the CMB temperature, 
Ref.~\cite{openhotteruniverse} notes that by varying the 
spatial curvature parameter $\Omega_K$ as well as $T_0$, tensions with the Hubble and other 
parameters can be reduced even when including BAO data, due to the extra freedom in the background evolution that curvature allows.  At roughly 10\,\% the required shift in $T_0$ is still much larger than the FIRAS uncertainty, although the authors of Ref.~\cite{openhotteruniverse} claim this can be explained by our presence in a large underdensity, 
which would render the locally measured $T_0$ colder than outside the void.  
However, as mentioned above the gravitational redshift or blueshift due to realistic structure is 
of order $10^{-5}$ over a large range of scales.  Thus the proposal 
of~\cite{openhotteruniverse} would require a potential well that is four orders of magnitude deeper than expected, likely conflicting with a range of observations~\cite{mzs10}.

\bibliography{tempdff}

\end{document}